\documentclass[12pt, oneside]{article}   	

\usepackage{jheppub}

\usepackage{amsmath}
\usepackage{amssymb}
\usepackage{amsfonts}
\usepackage{amsthm}
\usepackage{amsbsy}
\usepackage{array}
\usepackage{braket}
\usepackage[linesnumbered,lined,boxed,commentsnumbered]{algorithm2e}
\usepackage[OT2,OT1]{fontenc}
\newcommand\cyr
{
\renewcommand\rmdefault{wncyr}
\renewcommand\sfdefault{wncyss}
\renewcommand\encodingdefault{OT2}
\normalfont
\selectfont
}
\DeclareTextFontCommand{\textcyr}{\cyr}

\usepackage{mathtools}

\usepackage{tikz}
\usetikzlibrary{trees}
\usetikzlibrary{decorations.pathmorphing}
\usetikzlibrary{decorations.markings}
\usetikzlibrary{shapes.misc}
\usepackage{tkz-euclide}
\usetikzlibrary{arrows,positioning} 
\usepackage{tikzit}

\tikzset{
    >=stealth',
    punkt/.style={
           rectangle,
           rounded corners,
           draw=black, very thick,
           text width=15em,
           minimum height=2em,
           text centered},
    pil/.style={
           ->,
           thick,
           shorten <=2pt,
           shorten >=2pt,}
}

\usepackage{subcaption}

\usepackage{dsdshorthand}

\usepackage[vcentermath]{youngtab}

\tikzset{
  threept/.style={
    circle,
    draw,
    inner sep=2pt,
  },
  twopt/.style={
    circle,
    draw,
    fill=black,
    inner sep=1pt,
    minimum size=1pt
  },
  cross/.style={
    cross out,
    draw=black, 
    minimum size=7pt, 
    inner sep=0pt,
    outer sep=0pt
  },
  scalar/.style={
    thick,
    dashed,
    postaction={
      decorate,
      decoration={
        markings,
        mark=at position 0.5 with {\arrow{>}}
      }
    }
  },
  spinning/.style={
    thick,
    postaction={
      decorate,
      decoration={
        markings,
        mark=at position 0.5 with {\arrow{>}}
      }
    }
  },
  spinning no arrow/.style={
    thick,
  },
  finite with arrow/.style={
    decoration={
      snake,
      amplitude=1pt,
      segment length=6pt,
      post length=2pt
    },
    decorate,
    thick,->
  },
  finite/.style={
    decoration={
      snake,
      amplitude=1pt,
      segment length=6pt,
    },
    decorate,
    thick
  }
}

\newcommand\Sch{\text{Sch}}

\newcommand\mR{\mathbb{R}}

\newcommand\mC{\mathcal{C}}
\renewcommand{\tilde}[1]{\widetilde{#1}} % Like who does like this tiny tilde anyway %

\DeclareMathOperator{\diag}{diag}

\newcommand{\bea}{\begin{eqnarray}}
\newcommand{\eea}{\end{eqnarray}}
\def\ben{\begin{equation}}
\def\een{\end{equation}}
\def\bne{\begin{equation}}
\def\ene{\end{equation}}

\let\a=\alpha \let\b=\beta \let\g=\gamma \let\d=\delta \let\e=\varepsilon
   \let\k=\kappa
\let\l=\lambda \let\m=\mu    \let\r=v
\let\s=\sigma \let\t=\tau

\let\w=\omega \let\G=\Gamma \let\D=\Delta

\def\nn{\nonumber}

\def\ba{\begin{array}}
\def\ea{\end{array}}
\def\beq{\begin{equation}}
\def\eeq{\end{equation}}

\makeatletter
\def\@fpheader{\ }
\makeatother

\title{Classifying boundary conditions in JT gravity: from energy-branes to $\alpha$-branes}
\author[1]{Akash Goel,}
\affiliation[1]{Department of Physics, Princeton University, Princeton, NJ 08544, USA }
\author[1,2]{Luca V. Iliesiu,}
\affiliation[2]{Stanford Institute for Theoretical Physics, Stanford University, Stanford, CA 94305}
\author[2]{Jorrit Kruthoff,}
\author[2]{Zhenbin Yang}

\definecolor{jorrit-green}{RGB}{0, 131, 9}

\abstract{
We classify the possible boundary conditions in JT gravity and discuss their exact quantization. Each boundary condition that we study will reveal new features in JT gravity related to its matrix integral interpretation, its factorization properties and ensemble averaging interpretation, the definition of the theory at finite cutoff,  its relation to the physics of near-extremal black holes and, finally, its role as a two-dimensional model of cosmology. 
}
\date{}
\begin{document}

\maketitle

\addtolength{\parskip}{.5mm}

\newpage
\section{Introduction}
\label{sec:intro}

In recent years, two-dimensional Jackiw-Teitelboim (JT) gravity has emerged as an important toy model for quantum gravity and near-horizon physics \cite{Teitelboim:1983ux, Jackiw:1984je, kitaevTalks, Almheiri:2014cka, Jensen:2016pah,  Maldacena:2016hyu, Maldacena:2016upp, Engelsoy:2016xyb, Saad:2019lba}. The simplicity of this model allows for a detailed analytic analysis, which covers not only perturbative features of gravity \cite{Maldacena:2016upp}, but also non-perturbative ones, such as the sum over topologies \cite{Saad:2019lba}. However, most of these discussions focused on a particular set of boundary conditions that fix the asymptotic value of the dilaton and metric. Such boundary conditions can be taken to be part of the definition of the theory, and one might wonder whether there are other boundary conditions that one can impose and explore what new features they exhibit. 

The analysis of different boundary conditions in gravity is, of course, not new. In the past, it has been the subject of various studies \cite{PhysRevD.15.2752, PhysRevD.47.1420}. 
Nevertheless, the focus there was mostly on gravity theories in three or more space-time dimensions where the full quantization of the theory is not well understood. The purpose of this paper is to understand the role of boundary conditions in two-dimensional gravity and, in particular, in JT gravity, where the quantization of the theory is possible at the level of the path integral. The study of these different boundary conditions can reveal various aspects of the theory that are hard, or even impossible, to study in the (by now standard) Dirichlet boundary conditions for the metric and dilaton. Other discussions about boundary conditions in JT gravity can be found in \cite{Cvetic:2016eiv,Ferrari:2020yon}.

\begin{figure}
    \centering
    \includegraphics[width=\textwidth]{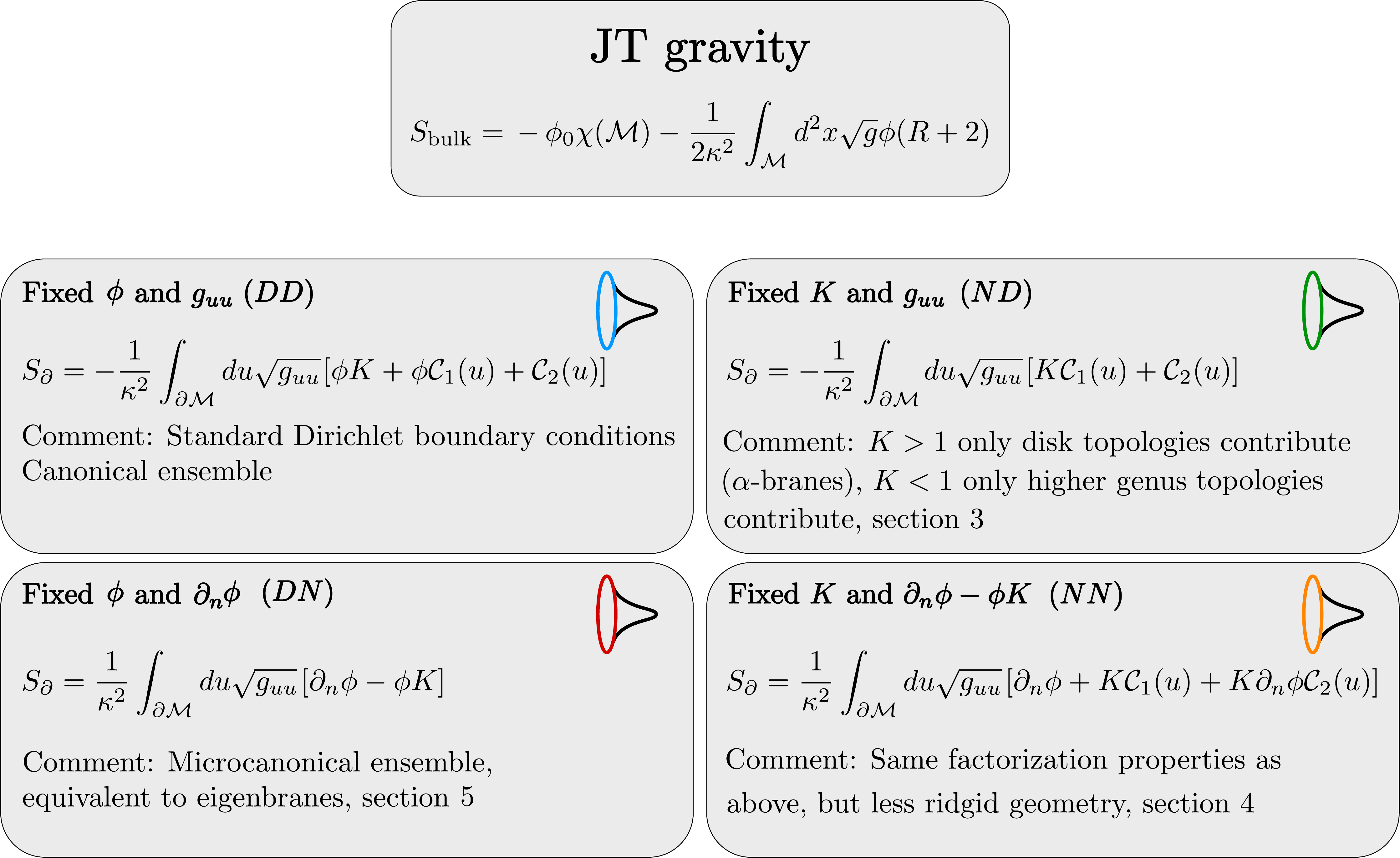}
    \caption{Summary of the boundary conditions considered in this paper. On the left column, one can see the boundary conditions for fixed dilaton $\phi$ (D) on the boundary, whereas the right column is for fixed $K$, i.e., Neumann boundary conditions on the dilaton (N). This is indicated by the first letter of the two letters in brackets. The rows are analogous, but then for the boundary metric $g_{uu}$ and is specified in the second letter in brackets.}
    \label{fig:summary}
\end{figure}

As we will explain, JT gravity admits four inequivalent boundary conditions obtained by fixing combinations between the dilaton $\phi$ and boundary metric $g_{uu}$ or their canonical momenta, schematically given by the extrinsic curvature $K$ and the derivative of the dilaton $\partial_n \phi$ normal to the boundary, respectively. The properties and features of these boundary conditions are summarized in figure \ref{fig:summary}.

 To exemplify and compare many of these features, we start by reviewing past results in the standard Dirichlet-Dirichlet (DD) boundary condition where the dilaton $\phi$ and the boundary metric $g_{uu}$ are fixed, when the cosmological constant in JT gravity is set to be negative. In the limit in which the value of the dilaton and the proper length of the boundary is large (the so-called Schwarzian limit), such boundary conditions correspond to the thermal canonical ensemble for the near-extremal black holes whose dynamics can be captured by JT gravity \cite{Maldacena:2016upp, Almheiri:2016fws, Nayak:2018qej, Moitra:2019bub, Sachdev:2019bjn,Ghosh:2019rcj, Iliesiu:2020qvm}. \footnote{Such black holes have their horizon located where the value of the dilaton (which captures the volume of the transverse space) is minimized. } Consequently, one might wonder what role other boundary conditions in JT gravity play when describing the near-horizon physics of near-extremal black holes. The other boundary condition whose meaning is clear fixes $\partial_n \phi$ instead of $g_{uu}$ (DN).  Such a boundary condition corresponds to the micro-canonical ensemble for the near-extremal black holes: instead of fixing the temperature of the system (which is the case when fixing $g_{uu}$) we fix the ADM mass of the black hole (which in dilaton gravity is given by $\partial_n \phi$). When setting $K$ instead of $\phi$ (ND or NN), the black hole interpretation is somewhat unclear. In such a case, we find that the dilaton solution cannot be fully fixed.\footnote{As we shall explain, the classical dilaton solution is fixed up to three undetermined constants. For DD or DN boundary conditions, these constants are fixed by the fact that $\phi$ is fixed on the boundary.  For ND and NN boundary conditions, all three constants cannot be fixed. } Therefore, the classical solutions do not exhibit the presence of a horizon and, consequently, such boundary conditions do not have any thermodynamical interpretation.

 When accounting for corrections from manifolds with other topologies, the partition function with the standard boundary conditions is reproduced by the insertion of the ``partition function'' operator $\Tr e^{-\beta H}$ in a specific double-scaled matrix integral which averages over the ``Hamiltonians'' $H$  \cite{Saad:2019lba, Saad:2018bqo, Stanford:2019vob}. Because of this interpretation as an ensemble average, or equivalently, due to the contribution from geometries that connect different boundaries, the partition function of the gravitational theory does not factorize when studying configurations with multiple boundaries. To improve the dictionary between gravity and its matrix integral ``dual'', we should determine what operator insertion on the matrix integral side corresponds to studying the three other previously mentioned boundary conditions in JT gravity. For instance, if instead of fixing $g_{uu}$, we fix $\partial_n \phi$, we find that the matrix operator insertion required for considering such boundaries is equivalent to the energy-eigenbranes (i.e. for $n$ boundaries, we fix $n$  eigenvalues of the random matrix $H$ to the ADM masses specified by $\partial_n \phi$ on each border) recently studied in \cite{Blommaert:2019wfy}. Once again, we find that the partition function of the theory with such boundary conditions does not factorize.  

Since both the DD and DN boundary conditions exhibit this feature, we can consequently ask whether there are other boundary conditions in JT gravity for which we instead have a factorizable multi-boundary partition function. As we shall explain shortly, when appropriately fixing $K$ (instead of the boundary value of the dilaton $\phi$), we will show that the multi-boundary partition function factorizes. Specifically, when setting $K>1$, we find that the multi-boundary partition function only receives contributions from geometries with disk topology and, therefore, factorizes.\footnote{When coupling the theory to matter (which we can take to couple solely to the metric, and not to the dilaton), our conclusions regarding the factorization of the multi-boundary partition functions for $|K|>1$ remains unchanged. In the context of black holes and in the presence of matter, the existence of Euclidean wormholes that connect different boundaries has been crucial to reproducing the Page curve predicted by requirements of unitarity \cite{Almheiri:2019qdq, Penington:2019kki, Almheiri:2020cfm}. Nonetheless, since the ND or NN boundary conditions do not have solutions that contain a clear horizon, the interpretation of the gravitational theory in terms of black hole physics is now unclear. } In the context of the baby universe Hilbert space discussed in \cite{Marolf:2020xie, Coleman:1988cy, Giddings:1988cx, Giddings:1988wv}, such boundaries can naturally be called $\alpha$-eigenbranes; i.e.~states that have a clear geometric meaning, such as the Hartle-Hawking state, are eigenfunctions of the operators corresponding to such boundary insertions. This should be contrasted with the case of the DD or DN boundary conditions for which the $\alpha$-states are an infinite linear combination of geometric states which can have an arbitrary number of boundaries. As opposed to $K>1$, for $K<1$, we find that the partition function receives no contribution from disk topologies, and only higher genus or connected multi-boundary geometries contribute.\footnote{Throughout this paper, we shall focus on the case with $K>0$. }

Another related direction that we study concerns the definition and quantization at finite cutoff. For the standard Dirichlet-Dirichlet boundary condition the quantization of the theory with finite proper length and dilaton value was studied in \cite{Iliesiu:2020zld, Stanford:2020qhm}. However, the exact quantization of the theory for all values of the proper length remains ambiguous: in \cite{Iliesiu:2020zld} non-perturbative corrections in the proper length to the partition function could not be fixed, while in \cite{Stanford:2020qhm} it was found that for small enough proper lengths, there is no effective theory which can describe the dynamics of the boundary. Furthermore, neither of these works fully accounted for the contributions from higher genus geometries.\footnote{However, we hope to report on progress on this direction in \cite{Iliesiu:toAppearTT}. } As we shall explain shortly, due to the rigidity of two-dimensional hyperbolic space, by fixing $K$ instead of $\phi$, the theory is easy to define at finite cutoff, and its partition function can be computed in a genus expansion. 

JT gravity, this time with a positive cosmological constant, has recently been used as a cosmological toy model \cite{Maldacena:2019cbz, Cotler:2019nbi, Chen:2020tes}. When computing expectation values of operators within this setting, one could use the path integral in the gravitational theory to prepare the density matrix of the system \cite{Maldacena:2019cbz,Penington:2019kki, Chen:2020tes,Anous:2020lka}.  Thus, one is typically interested in computing the path integral for both connected and disconnected geometries in the presence of two or more boundaries. If one fixes the standard Dirichlet-Dirichlet boundary condition, the contribution of connected geometries, some of which are called bra-ket wormholes, resolves an inconsistency in the entropy of the theory when coupled to matter \cite{Penington:2019kki,Chen:2020tes}. Since the connected geometries play such an important role, one can ask, just like in the case of AdS$_2$, whether there exist boundary conditions which disallow the existence of such connected manifolds with multiple boundaries. Once again, we find that by appropriately fixing $K$ (this time, $K<1$) the path integral which prepares the density matrix in the gravitational theory does not receive contributions from connected geometries, and, in particular, from the bra-ket wormhole geometries. Such boundary conditions, where one fixes $g_{uu}$ in addition to $K$, are in fact, quite natural in higher-dimensional models of cosmology \cite{PhysRevLett.28.1082}; in such a case, one fixes the trace of the extrinsic curvature $K$ and the conformal metric. $K$ can then be taken to be a clock (called York time) that parametrises the slices in the bulk.

  The rest of this paper is organized as follows. In section \ref{sec:class-bdy-cond} we present the full classification of boundary conditions in theories of dilaton gravity and derive the necessary boundary-terms and possible counter-terms consistent with the variational principle. In section \ref{sec:Kguu}, we derive the classical solutions when fixing $g_{uu}$ and $K$ and study the quantization of the theory from various perspectives. We use the same techniques to study the fixed $\partial_n \phi - \phi K$ and $K$ boundary conditions in section \ref{sec:NN-bc} and the fixed $\partial_n \phi$ and $\phi$ boundary conditions in section \ref{sec:DN-bc-eigenbranes}. Finally, in section \ref{sec:discussion}, we discuss the features of the novel boundary conditions in the context of black hole physics and cosmology and rephrase some of the results obtained in the previous sections in the language of the baby universe Hilbert space, recently discussed in \cite{Marolf:2020xie}. Furthermore, in this section, we analyze our results in the context of AdS/CFT and analyze the relationship between the various b.c.~that we impose in JT gravity and b.c.~that can be imposed in the $(2,p)$ minimal string.

\section{A classification of boundary conditions} 
\label{sec:class-bdy-cond}

Before diving into the different boundary conditions central to the study in this paper, let us first review briefly the standard case, discussed in \cite{Maldacena:2016upp}, so that we are all on the same page. The standard case amounts to putting Dirichlet boundary conditions on both the metric and dilaton at asymptotic infinity. In Euclidean signature it is convenient to think of fixing the metric as fixing the proper boundary length. 

Let us consider Euclidean JT gravity on a manifold $\cM$ and boundary $\partial \cM$ with action\footnote{The Lorentzian theory can be considered analogously.  We will study both the Euclidean and Lorentzian theories for the boundary conditions encountered later on.} 
\be 
\label{eq:JT-gravity-topo+bulk+bdy}
S = S_{\rm top} + S_{\rm bulk} + S_{\partial}.
\ee
The first of these three terms is given by $S_{\rm top} = -\frac{\phi_0}{4\pi}\left(\int_{\cM}\sqrt{g} R + 2 \int_{\partial \cM}  \sqrt{h} K \right) =  -\phi_0 \chi(\cM)$ with $\phi_0$ a constant and $\chi(\cM)$ the Euler character of $\cM$. The other two pieces are given by
\be\label{Sbulkbndry}
S_{\rm bulk} = - \frac{1}{16\pi G_N}\int_{\cM} d^2 x \sqrt{g}\,\phi(R + 2), \quad S_{\partial} = - \frac{1}{8\pi G_N} \int_{\partial \cM} du \sqrt{h}\,\phi(K-1),
\ee
where henceforth we work in units where $16\pi G = 1$, $g$ is the bulk $2d$ metric, $\phi$ the dilaton and $h$ and $K$ are the induced metric and extrinsic curvature on the boundary $\partial \cM$. The boundary term here is chosen so that the variational principle is well-defined ($K$ dependent term) and make the on-shell action finite (second term). It is this term that we will modify momentarily, but let us first continue discussing the standard Dirichlet case. 

The equations of motion for the metric and dilaton are given by 
\be\label{eom}
R + 2 = 0,\quad \left(\nabla_\mu \nabla_\nu - g_{\mu\nu}\nabla^2 + g_{\mu\nu}\right) \phi = 0.
\ee
The metric is thus that of Euclidean AdS$_2$ ($\mathbb{H}^2$). The on-shell solution of the dilaton is parameterized by a three dimensional vector $Z$ in embedding space corresponding to the different $SL(2,\mathbb{R})$ charges of the dilaton field. Denoting the $\mathbb{H}^2$ embedding coordinates by $Y$ we have $-Y_0^2 + Y_1^2 + Y_2^2 = -1$ and $\phi=Z.Y$.
In particular, with Poincar\'{e} coordinates the metric and dilaton profile are given by
\be \label{CSoln}
ds^2 = \frac{d\t^2 + dz^2}{z^2},\quad \phi = \frac{a_1 + a_2 \t + a_3(\t^2+z^2)}{z}.
\ee
The boundary is located at $z \to 0$, but we can consider cutting the geometry along some curve $(\t(u),z(u))$, with $u$ the intrinsic boundary coordinate. The boundary conditions we then impose \cite{Maldacena:2016upp} on the curve $(\t(u),z(u))$ are
\be 
ds^2|_{\partial \cM} = g_{uu}du^2 = \frac{du^2}{\e^2},\quad \phi|_{\partial \cM} = \frac{\phi_r(u)}{\e}
\ee
with $\e > 0$ small. We can solve these boundary conditions in an expansion in $\e$ and to leading order in $\e$ we find, $
z(u) = \e \tau'(u)$ and a corresponding solution for the dilaton.  The boundary is thus at asymptotic infinite, close to $z = 0$ as $\e \to 0$. For an outward pointing normal vector $n^\mu$ to the boundary, the extrinsic curvature is defined as $K = \nabla_{\mu}n^\mu$, and in this case given by, 
\be
K = 1 +\e^2 \Sch(\tau, u)\,, \qquad \Sch(\tau, u) = \frac{\tau'''}{\tau'} - \frac{3}2\left(\frac{\tau''}{\tau'}\right)^2
\ee
where $\tau' = \partial_u \tau$ and $\Sch(\tau, u)$ is the Schwarzian derivative. 
Since the bulk term in \eqref{eq:JT-gravity-topo+bulk+bdy} vanishes after integrating out the dilaton $\phi$, the only remaining degree of freedom is given by the reparametrization mode $\tau(u)$, whose action is given by the Schwarzian derivative defined above. Consequently, the quantization of the theory (for disk topologies) solely reduces to the quantization of the Schwarizan theory. One can also use the quantization of the Schwarizan theory together with results for the volumes of the moduli space of constant curvatures Riemann surfaces to compute the contribution of surfaces with any topology to the JT partition function.   The quantization of this theory was studied in great detail in \cite{Mertens:2017mtv, Kitaev:2018wpr, Yang:2018gdb, Mertens:2018fds,Saad:2019lba,   Iliesiu:2019xuh} and the contribution of higher genus surfaces was computed in \cite{Saad:2019lba}. We will not further review the results for the Dirichlet-Dirichlet boundary conditions here, and, instead, we will now consider JT gravity with different boundary conditions. For each case, we will conduct both a classical and quantum analysis similar to the Dirichlet-Dirichlet case, briefly discussed above. However, as opposed to the Dirichlet case, for several boundary conditions that we study, we will easily be able to go beyond the nearly-AdS$_2$ limit, to a finite patch of AdS$_2$.

To start our analysis, we first review the variation of the bulk action (see also appendix \ref{app:variationbulkaction} for details) and map out all different boundary conditions consistent with a well-defined variational problem together with the appropriate boundary terms. We point out that there are two different sets of conjugate variables that are natural to consider, giving, in total 6 different type of boundary conditions, of which only $4$ are physically inequivalent. In subsequent sections we will then discuss these boundary conditions (excluding the standard Dirichlet case) in more detail, not only classically, but also quantum mechanically.  

The bulk action for JT gravity was given in \eqref{Sbulkbndry} and in appendix \ref{app:variationbulkaction} we review its variation. The variation of this bulk term is given by,
\be
\delta S_\text{bulk} =-\text{EOM} -\int_{\partial \cM} du \sqrt{h} \left( 
\left[(\partial_n\phi)h^{\mu\nu} - \phi K^{\mu\nu} \right]\delta h_{\mu\nu} - 2 \phi\delta K
\right)\,,
\ee
where $\text{EOM}$ refers to the equations of motion in \eqref{eom}, $h_{\mu \nu}$ is the induced boundary metric and $K$ is the trace of the extrinsic curvature. We can reduce to two dimensions $K^{\mu\nu} = K h^{\mu\nu}$ and express the boundary metric in terms of the variable $g_{uu}$
\be 
g_{uu} = h_{\mu\nu} \frac{\partial x^\mu}{\partial u}\frac{\partial x^\nu}{\partial u} \qquad \Leftrightarrow\qquad ds^2 = h_{\mu\nu} dx^{\mu} dx^{\nu} = g_{uu} du^2\,,
\ee
This results in 
\be 
\label{sbulkvar}
\delta S_{bulk} = -\text{EOM} - \int_{\partial \cM} du \left( \left[ 2 \left(\partial_n\phi - \phi K \right) \right] \delta \left(\sqrt{g_{uu}}\right) - \left[ 2 \phi \sqrt{g_{uu}} \right]\delta K \right)\,.
\ee
From this we immediately recognize the following canonically conjugate pairs
\be 
\label{eq:canonical-pairs-1}
 \phi \sqrt{g_{uu}} &\leftrightarrow K\,,\nn \\
\sqrt{g_{uu}} &\leftrightarrow \partial_n\phi - \phi K \,, 
\ee
Our convention is to denote the left side of the pairs in \eqref{eq:canonical-pairs-1} to be associated to the coordinates while the right side are the canonical momenta. Fixing the coordinate means a Dirichlet boundary condition, which we will abbreviate with $D$, whereas for fixing the canonical momentum, we will use $N$ to denote a Neumann boundary condition. As we will discuss below the choice of canonical conjugates pairs is not unique and it is interesting to study the spectrum of choices.  

At the level of the path integral it is natural to study the case in which one fixes on the boundary either one of the two sides of the two canonical pairs in \eqref{eq:canonical-pairs-1}. This allows us to study the following four possible boundary conditions and the boundary terms required in order for the problem to have a well defined variational principle:
\vspace{0.5cm}

\subsubsection*{\it DD : Fixed $\phi$ and $g_{uu}$}
This is the standard JT gravity theory with Dirichlet boundary conditions. In this case, we require the addition of a boundary term \footnote{Below, when writing $S_{DD/ND/DN/NN}$ we are not including the topological term in \eqref{eq:JT-gravity-topo+bulk+bdy}. The variational problem for this term will be discussed at the end of this section.  }
\be
S_{DD} = S_\text{bulk} - 2\int du\, \sqrt{g_{uu}} \left[\phi K + \mC_1(u) \phi +\mC_2(u) \right]\,.
\ee
Here, $\mC_1(u)$ and $\mC_2(u)$ are arbitrarily fixed functions that do not affect the equation of motion. Thus, they serve as possible counter-terms to be added to the Euclidean path integral. In the usual JT gravity literature it is customary to take these to be $\mC_1(u) =-1$ and $\mC_2(u) = 0$ which fixes the vacuum energy of the equivalent boundary Schwarzian theory to be $0$ and renders the on-shell action finite. 
\vspace{0.5cm}
 
\subsubsection*{\it ND : Fixed $K$ and $g_{uu}$} 
In this case one does not require an additional boundary terms. possible counterterms which do not affect the equations of motion are given by  
\be
S_{ND} = S_\text{bulk} - 2\int du\, \sqrt{g_{uu}} \left( K \mC_1(u) + \mC_2(u)\right)\,.
\ee
When both $\mC_1(u)=0$ and $\mC_2(u)=0$ the equivalent $\mathfrak{sl}(2, \mR)$ gauge theory is purely topological.
\vspace{0.5cm}

\subsubsection*{\it NN$\,$ : Fixed $K$ and $ \left(\partial_n\phi - \phi K \right)$} 
In this case the additional boundary term is 
\be
\label{eq:NN-action}
S_{NN} = S_\text{bulk} +2 \int du\, \sqrt{g_{uu}} \left(\partial_n\phi -\phi K\right)\,.
\ee
If we again specialise to $ \left(\partial_n\phi - \phi K \right)$ constant, the boundary action is proportional to the length of the boundary. However, since we are not fixing the boundary metric $\sqrt{g_{uu}}$,  if we do choose $ \left(\partial_n\phi - \phi K \right)$ and $K(u)$ to not be a constant (but rather depend on $u$) we have to specify how to choose the parametrization of the boundary time. One can for instance choose to fix a boundary diffeomorphism gauge in which $\sqrt{g_{uu}} = \text{constant}$ and $u\to [0, 2\pi]$, without fixing the diffeomorphism invariant proper length. In such a case the boundary observer specifies the $2\pi$ periodic functions $ \left(\partial_n\phi - \phi K \right)$ and $K(u)$.\footnote{Alternatively, if we do not fix a boundary diffeomorphism gauge we can specify the values of $K(u)$ and $ \left(\partial_n\phi - \phi K \right)(u)$ according to the boundary proper length from a given boundary point. While this makes sense in Lorentzian signature, in Euclidean signature where the boundary is periodic, the proper boundary length is no longer fixed and therefore we do not know the periodicity of the functions $K(u)$ and $ \left(\partial_n\phi - \phi K \right)(u)$.  Furthermore, there is no consistent diffeomorphism invariant counter-term that could be added to the action.}
\vspace{0.5cm}
 
\subsubsection*{\it DN$\,^*$ : Fixed $\phi\sqrt{g_{uu}}$ and $\left(\partial_n\phi - \phi K \right)$} 
In this case, both the boundary terms need to be added 
\be 
S_{DN}  = S_\text{bulk} +2 \int \sqrt{g_{uu}} \left(\partial_n\phi - 2 \phi K + \phi \,\mC_1(u) + (\partial_n\phi - \phi K)\phi \,\mC_2(u) \right)\,.
\ee
where $\mC_1(u)$ and $\mC_2(u)$ represent the possible counter-term whose variation vanishes. 
\vspace{1.5cm}

An alternate set of conjugate variables can be obtained by rewriting \eqref{sbulkvar} as, 
\be
\label{eq:variation-alternative}
\delta S_\text{bulk} = \int du \left( \left[ 2 \phi  \right]\delta\left(\sqrt{g_{uu}} K \right) -\left[ 2 \left(\partial_n\phi \right) \right] \delta \left(\sqrt{g_{uu}}\right)  \right) \,.
\ee
With this choice, the pair of canonical conjugates pairs are given by
\be 
 \phi &\leftrightarrow \sqrt{g_{uu}}  K\,,\nn \\
\sqrt{g_{uu}} &\leftrightarrow \partial_n\phi  \,.
\ee
Using this pair of canoncial variables two alternative boundary conditions follow: 
\subsubsection*{\it NN$^*$: Fixed $\sqrt{g_{uu}}K$ and $ \left(\partial_n\phi\right)$}
In this case the additional boundary term is 
\be
S_{NN^*} = S_{bulk} +2 \int du\, \sqrt{g_{uu}}\, \left[\partial_n\phi + \mC_1(u) K + ( K \partial_n\phi ) \mC_2(u)\right]\,.
\ee
Again, $\mC_1(u)$ and $\mC_2(u)$ represent the possible counter-terms. 
\vspace{0.5cm}

\subsubsection*{\it DN: Fixed $\phi$ and $\left(\partial_n\phi\right)$} 
In this case, two boundary terms need to be added in order to cancel the variation in \eqref{eq:variation-alternative}
\be
S_{DN^*} 
 = S_{bulk} +2 \int du \sqrt{g_{uu}} \left(\partial_n\phi - \phi K\right)\,.
\ee
As in the case of NN boundary conditions there are no possible counter-terms which are invariant under boundary diffeomorphisms. 
\vspace{1.2cm}

We can also analyze the variation of the topological term in \eqref{eq:JT-gravity-topo+bulk+bdy}. On the one hand, because the bulk plus boundary terms in the action is invariant under metric fluctuations (since the action solely depends on the Euler characteristic of the manifold), it is clear that the variation should vanish regardless of the boundary conditions that we imposed. On the other hand, explicitly checking that the boundary term of the topological term arises from consistency with the variational principle serves as a non-trivial consistency check. Following \ref{sbulkvar}, the variation of the bulk topological term is 
\be 
\label{topbulkvar}
\delta S_{\text{top, \,bulk}} = \int_{\partial \cM} du \left[  2   \phi_0 K  \delta \left(\sqrt{g_{uu}}\right) +  2 \phi_0 \sqrt{g_{uu}} \delta K \right]\,.
\ee
With the addition of the boundary topological term the variation of $S_\text{top}$ always vanishes regardless of whether we set $\delta K = 0$, $\delta \sqrt{g_{uu}} = 0$ or $\delta( \sqrt{g_{uu}} K) = 0$.

In summary, we have listed $6$ boundary conditions for JT gravity, which, in fact are also valid for any other dilaton gravity. However, since we only have two pairs of conjugate variables, it suffices to study four of them or three actually as we will not consider the standard Dirichlet case here. Below we will therefore consider the boundary conditions ND, NN and DN in more detail. We will study the associated classical solutions space, the quantum theory, coupling to matter and comment on the physical aspects of the boundary conditions and what properties of JT gravity they probe.

\section{Fixing $K$ and $g_{uu}$: towards factorization }
\label{sec:Kguu}

 The fixed $K$ boundary conditions have a less clear interpretation from the ``boundary dual'' point of view.\footnote{Such boundary conditions are however very natural to consider in the context of cosmology, where one can take $K$ to be a clock (called York time) that parametrises slices in the bulk. In the current context, however, we have a time-like boundary and we cannot use this intuition.} Still, JT gravity with these boundary conditions is a rich theory that allows us, as we will show, to turn on (leading to factorization) or off (leading to an ensemble average) the sum over topologies depending on  the value of the extrinsic curvature. As we shall explain below, in the particular case when $K= 0$, the partition function of JT gravity is given by the Weil-Peterson volumes whose recursion relation was determined by Mirzakhani \cite{Mirzakhani:2006fta, Mirzakhani:2006eta}. Thus, the results for the partition function with boundaries given by different values of $K$ and $g_{uu}$ can be viewed as a generalization of the results previously obtained in the mathematical literature. 

\subsection{Classics}

Before we begin our classical analysis, we emphasize that, as opposed to the Dirichlet boundary conditions reviewed in section \ref{sec:class-bdy-cond}, or to the micro-canonical boundary conditions which we will study in section \ref{sec:DN-bc-eigenbranes} \cite{Saad:2019lba}, the fixed  $K$ and $g_{uu}$ boundary conditions allow for the existence of richer on-shell higher topology or multi-boundary solutions.\footnote{This is because in the previously studied cases the problem with the existence of such solutions stemmed from imposing a boundary value of the dilaton and from imposing the dilaton equation of motion in the bulk. In this section, since we do not fix the boundary value of the dilaton, we can always study the trivial (but not necessarily unique) solution $\phi(x) = 0$. }
 
 With this in mind, we start by studying solutions when $K \equiv k$ is fixed to a constant. To understand such solutions we provide three different perspectives. The first is purely geometric and reviews the classification of constant $K$ curves on the Poincar\'e plane or disk and on its quotients. The second explicitly solves for fixed $K$ curves by working in the Fefferman-Graham gauge for the metric. Using this metric, it becomes clear why fixing $|k|>1$ solely isolates disk topologies while $|k|<1$ isolates topologies with a higher genus or a higher number of boundaries. Finally, in our third  
approach we will again solve for curves of constant $K$, this time reformulating JT gravity as a $PSL(2, \mR)$ BF theory. This latter perspective allows us to study the case with varying $K \equiv k(u)$. In particular, we show that such solutions can in turn be mapped to solutions of the Hill equation whose classification has been extensively studied.

\subsubsection{A geometric construction}

\begin{figure}
\begin{center}
 \begin{subfigure}[b]{0.45\linewidth}
\begin{tikzpicture}[scale=3.5]
    \tkzDefPoint(-1,0){A1}
    \tkzDefPoint(0,0){O}
    \tkzDefPoint(-0.75,0.3){O2}
    \tkzDefPoint(-0.75,0){A4}
    \tkzDefPoint(0.5, 0){A3}
    \tkzDefPoint(1,0){A2}
    \tkzDefPoint(0.5,0.85){O5}
    \tkzDefPoint(0.5,0.4){A5}
    \tkzDefPoint(0,0.3){O6}
     \tkzDrawCircle[color=red](O,A3)
     \tkzDrawCircle[color=jorrit-green](O2,A4)
     \tkzDrawCircle[color=jorrit-green](O5,A5)
     \tkzDrawCircle[color=jorrit-green](O6,A3)
    \tkzText(0,0.32){$k = 0$}
    \tkzText(-0.75,0.65){$k = 1$}    
     \tkzText(-0.5,0.85){$|k| < 1$}  
      \tkzText(-0.05,1.2){$|k| > 1$}  
    \draw (A1) -- (A2);
    
\end{tikzpicture}
\end{subfigure}
 \begin{subfigure}[b]{0.45\linewidth}
\begin{tikzpicture}[scale=3.5]
  \tkzDefPoint(0,0){O}
  \tkzDefPoint(1,0){A}
  \tkzDrawCircle(O,A)
  \tkzDefPoint(0.5,-0.25){z1}
  \tkzDefPoint(-0.35,-0.5){z2}
  %\tkzDrawArc[color=blue](O,A)(z1)
  \tkzDefPoint(cos(pi/2*50/90), -sin(pi/2*50/90)){z3}
  \tkzDefPoint(cos(pi/2*105/90), -sin(pi/2*105/90)){z4}
  \tkzClipCircle(O,A)
   
  \tkzDrawCircle[orthogonal through=z1 and z2, color=red](O,A)
   \tkzGetPoint{O1} 
 %  \tkzDrawArc[color=blue](O1,z1)(z2)
 % \tkzDrawCircle[orthogonal through=z1 and z3, color=red, style=dashed](O,A)
    \tkzGetPoint{O2} 
%   \tkzDrawCircle[orthogonal through=z2 and z4, color=red, style=dashed](O,A)
       \tkzGetPoint{O3} 
  %\tkzDrawPoints[color=black,fill=red,size=12](z1,z2)
 \tkzInterCC(O,A)(O1,z1)\tkzGetPoints{C}{D}
 \tkzDefPointBy[homothety = center O ratio .5](O1) \tkzGetPoint{centerHyper}
  \tkzDrawCircle[color=jorrit-green](centerHyper,D)
   \tkzDefPointBy[homothety = center O ratio -1.2](O1) \tkzGetPoint{centerHyper}
   
    \tkzDefPoint(0.6*cos(pi/2*50/90), 0.6*sin(pi/2*50/90)){OH}
    \tkzDefPoint(cos(pi/2*50/90), sin(pi/2*50/90)){AH}
    \tkzDrawCircle[color=jorrit-green](OH,AH)
    
        \tkzDefPoint(-0.6*cos(pi/2*50/90), 0.6*sin(pi/2*50/90)){OC}
    \tkzDefPoint(-0.83*cos(pi/2*50/90), 0.83*sin(pi/2*50/90)){AC}
    \tkzDrawCircle[color=jorrit-green](OC,AC)
 % \tkzDrawCircle[color=purple](centerHyper,D)
  \tkzText(-0.08,-0.2){$|k| \leq 1$}
  \tkzText(0.1,-0.45){$k=0$}
   \tkzText(-0.38, 0.45){$|k|> 1$}
   \tkzText(0.38, 0.45){$k=1$}
    % \tkzText(0.8, 0.1){$G_1$}
    % \tkzText(-0.75,-0.4){$G_2$}
    \tkzDefPointBy[rotation= center O2 angle -85](z3)
%\tkzGetPoint{zCircle2}
%     \tkzDrawArc[color=red](O2,zCircle2)(z3)
 %          \tkzDefPointBy[rotation= center O3 angle 90](z4)
%\tkzGetPoint{zCircle3}
    % \tkzDrawArc[color=red](O3,z4)(zCircle3)
  %\tkzDrawPoints[color=black,fill=red,size=12]( O2, z3, zCircle2)
\end{tikzpicture}
\end{subfigure}
\end{center}
\caption{\label{fig:curve-cont-k} Figure showing the different curves of constant $k$ in the Poincar\'e plane and disk. The sign of $k$ is dependent on the direction of the normal vector perpendicular to the boundary; for instance, for $|k|>1$, $k>1$ is the boundary of the compact space (the disk), and $k<-1$ is the boundary of the non-compact space (the outside of the disk). In this paper, we will focus on $k>0$ to simplify our higher genus analysis.   }
\end{figure}
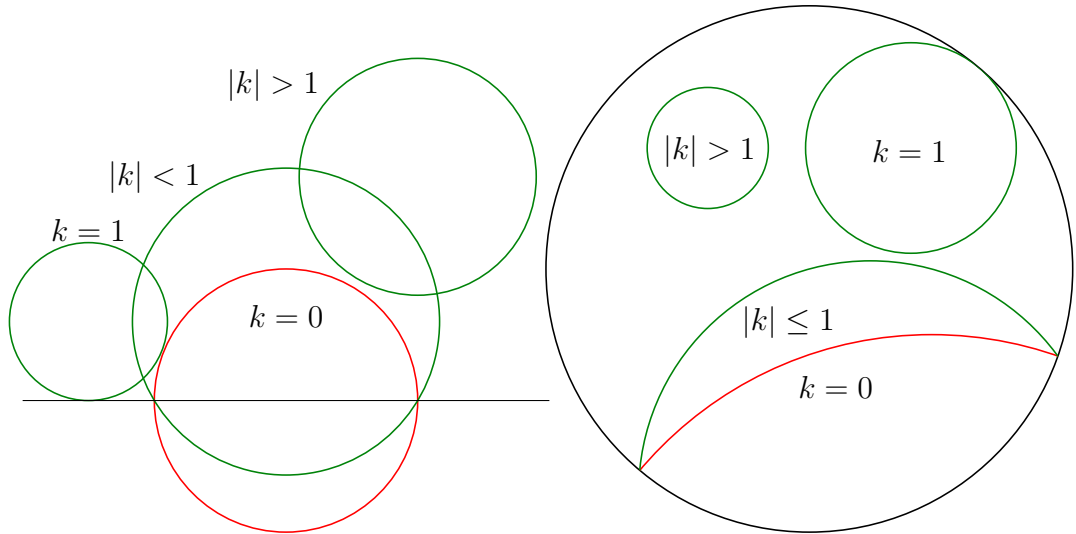

Before explicitly solving for curves of constant $K = k$ on the hyperbolic plane, it is useful to develop a geometric intuition that will be made concrete through the explicit computations in the next subsection. Just as in flatspace, curves with constant $k$ are circles in the Poincar\'e half-plane or disk. The circles in the Poincar\'e half-plane can be classified by their relation to the axis that bounds the half-plane (or on the Poincar\'e disk by their relation to the boundary of the disk):
\begin{itemize}
    \item \textbf{Geodesics }$(k=0)$. As is well known the geodesics of hyperbolic space are the semi-circles perpendicular to the axis (or the boundary circle of the Poincar\'e disk).
    \item \textbf{Hypercycles }$(|k|<1)$ are the circles that  intersect the boundary at exactly two points. An alternative definition that we will make use of is that it is the locus whose points have the same orthogonal distance from a given geodesic.   
    \item \textbf{Horocycles} ($|k| = 1$) are the circles which are tangent to the boundary of the Poincar\'e half-plane or disk.
    \item \textbf{Curves of interest.} ($|k| > 1$) are the circles which are fully contained in the Poincar\'e half-plane or in the disk.
\end{itemize}
We show examples of such curves in figure \ref{fig:curve-cont-k}. We will for now focus on $k \geq 0 $. Since curves with $k<1$ always intersect the boundary of hyperbolic space we can therefore never have a disk geometry whose boundary has $k<1$ at every point. Therefore,  solutions with disk topology only exist for $k>1$.

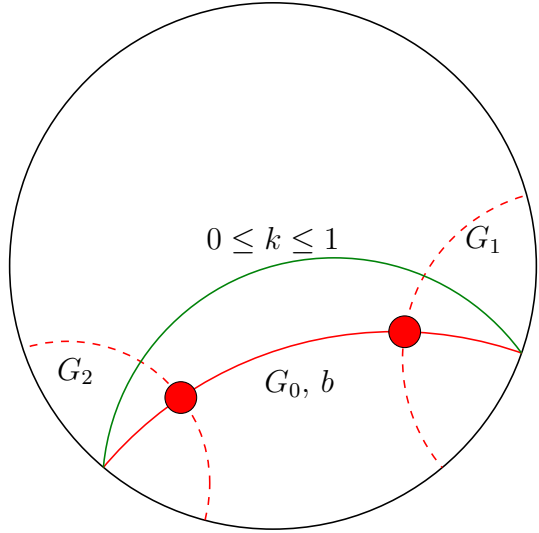
\begin{figure}
\begin{center}
\begin{tikzpicture}[scale=3.5]
  \tkzDefPoint(0,0){O}
  \tkzDefPoint(1,0){A}
  
  \tkzDrawCircle(O,A)
  \tkzDefPoint(0.5,-0.25){z1}
  \tkzDefPoint(-0.35,-0.5){z2}
  \tkzDefPoint(cos(pi/2*50/90), -sin(pi/2*50/90)){z3}
  \tkzDefPoint(cos(pi/2*105/90), -sin(pi/2*105/90)){z4}
  \tkzClipCircle(O,A)
  \tkzDrawCircle[orthogonal through=z1 and z2, color=red](O,A)
   \tkzGetPoint{O1} 
  \tkzDrawCircle[orthogonal through=z1 and z3, color=red, style=dashed](O,A)
    \tkzGetPoint{O2} 
   \tkzDrawCircle[orthogonal through=z2 and z4, color=red, style=dashed](O,A)
       \tkzGetPoint{O3} 
  \tkzDrawPoints[color=black,fill=red,size=12](z1,z2)
 \tkzInterCC(O,A)(O1,z1)\tkzGetPoints{C}{D}
 \tkzDefPointBy[homothety = center O ratio .5](O1) \tkzGetPoint{centerHyper}
  \tkzDrawCircle[color=jorrit-green](centerHyper,D)
   \tkzDefPointBy[homothety = center O ratio -1.2](O1) \tkzGetPoint{centerHyper}
  \tkzText(0,0.1){$0\leq k \leq 1$}
    \tkzText(0.1,-0.45){$G_0, \,b$}
    \tkzText(0.8, 0.1){$G_1$}
    \tkzText(-0.75,-0.4){$G_2$}
    \tkzDefPointBy[rotation= center O2 angle -85](z3)
\end{tikzpicture}
\end{center}
\caption{\label{fig:hyperbolic-quotient} Figure showing a region of a trumpet which ends on a single geodesic boundary $G_0$ of length $b$. The red curves represent geodesics on the Poincar\'e disk and the two geodesics $G_1$ and $G_2$ which are dotted are identified ($G_1 = G_2$). In order to be able to identify such two geodesics we require them to be perpendicular to the closed geodesic containing the segment of length $b$. To construct boundaries with constant $K$ in the same homotopy class as the geodesic boundary with length $b$ we construct hypercycles (shown in green) which, by definition, intersect the two geodesics $G_1$ and $G_2$ at equal distances. Note that because the hypercycles need to pass through the points where $G_0$ intersects the boundary, the hypercycles  can never be fully contained on the Poincar\'e disk and always have $k < 1$.   }
\end{figure}

We now explain why a stronger statement also follows from the above classification: when $0\leq k<1$, solely higher topology or multi-boundary geometries contribute, while if $k>1$ only the disk topology contributes. To do this, we first note that in order to construct higher genus or multi-boundary surfaces we need to glue a bordered higher genus Riemann surface whose boundaries are all closed geodesics to ``trumpets''. Such ``trumpets'' on one side have a closed geodesic, while on the other they have the fixed $K$ and $g_{uu}$ boundary that we are interested in. Thus, in order to understand the higher genera or multi-boundary case we need to study the possible boundaries that can be present on one end of the trumpet. To construct the trumpet, we consider the $\mathbb Z_2$ quotient of hyperbolic space by identifying two geodesics (for instance $G_1$ and $G_2$ in Figure \ref{fig:hyperbolic-quotient}). Then, the trumpet is identified as the (top) patch separated by the three geodesics $G_0$, $G_1$, and $G_2$, where $G_0$ is perpendicular to both $G_1$ and $G_2$. Thus, on the trumpet, $G_0$ is the closed geodesic whose length, $b$, is the distance between the geodesics $G_1$ and $G_2$. To construct the other boundary of the trumpet, we need to consider a fixed $K=k$ curve that intersects $G_1$ and $G_2$ at an equal distance from the intersection points between these two geodesics and $G_0$ (the red points in figure \ref{fig:hyperbolic-quotient}). Furthermore, in order for the fixed $K=k$ curve to be smooth when taking the $\mathbb Z_2$ quotient, the curve of fixed $K=k$ should be perpendicular to the two geodesics. Consequently, we are precisely replicating the definition of a hypercycle which we have reviewed earlier, where this hypercycle has a constant distance from the geodesic $G_0$. Therefore, only hypercycles (with $k<1$) can serve as the other boundary of the trumpet and, consequently, higher genus or multi-boundary surfaces can all only have boundaries with $k<1$. Next, we explicitly solve for such solutions and show that fixing $k$ uniquely fixes the proper length of the boundary.

\subsubsection{Finding the boundary curve}

Let us focus on Euclidean signature. To fix $K= k$ in the usual Poincar\'e metric where we cut out an arbitary shape parametrized by $(t(u),z(u))$ is non-trivial, since the expression for $K$ is rather cumbersome. We would therefore like to study the fixed $K$ boundary conditions with a different metric. A convenient metric, which has a more manageable expression for $K$ and that solves  $R + 2 = 0$ is one in Fefferman-Graham gauge
\be\label{eq:metricFG}
ds^2 = \frac{dr^2}{r^2} + \left(r - \frac{b(u)}{r}\right)^2 du^2.
\ee
This is a solution for any (smooth) $b(u)$. We will normalise the coordinate $u$ such that $u \sim u + 2\pi$ and hence, $b(u)$ needs to be $2\pi$ periodic. However it is important to note that not every $b(u)$ gives rise to a smooth geometry. 

For this metric, the extrinsic curvature on constant $r$ slices and their length is,
\be 
K(u) = \nabla_{\mu}n^{\mu} \equiv k(u) =  \frac{1 + \tilde{K}(u)}{1 - \tilde{K}(u)}, \quad \tilde{K}(u) = \frac{b(u)}{r^2},\quad L = \int_0^{2\pi} du\, \left| r - \frac{b(u)}{r} \right| 
\ee
where we  have defined $\tilde{K}$ for later convenience and $n^\mu$ is an outward pointing normal vector to the boundary. From this, the advantage of using Fefferman-Graham gauge for these boundary conditions is also clear, since $K(u)$ is so simple. In order to gain some intuition for this metric, let us again first focus on $u$-independent metrics: $k(u) = k$ and so $b(u) \equiv b_0$.\footnote{In embedding coordinates $Y_i$ the geometry is given by 
\be
Y_0 = \frac{\sqrt{b_0}}{2r}\left(1 + \frac{r^2}{b_0}\right),\quad Y_1 = \frac{1}{2\sqrt{b_0}}\left(r - \frac{b_0}{r}\right)\cos(2\sqrt{b_0}u),\quad Y_2 = \frac{1}{2\sqrt{b_0}}\left(r - \frac{b_0}{r}\right)\sin(2\sqrt{b_0}u).
\ee
} 
The time-dependent solutions are more intricate to consider and will be done below. 

For $b_0 = 0$ this metric is the Poincar\'{e} patch of AdS$_2$, with (non-compact) boundary at $r = \infty$ and the Poincar\'{e} horizon at $r = 0$. When $b_0 > 0$ the metric has a conical singularity at $r = \sqrt{b_0}$. This singularity is avoided when we pick $b_0 = 1/4$. For a smooth geometry, the value of $b_0$ is thus fixed,\footnote{If we allow a different periodicity for $u$, say $u \sim u + \b$, $b_0$ will be fixed to $\pi^2/\b^2$.} and take $r > 1/2$ as our space-time with the boundary located at $r = \infty$.\footnote{One could also take $r < 1/2$, but that patch can be mapped to the $r > 1/2$ patch through the diffeomorphism $r \to \frac{1}{4r}$.} 

In the case of $b_0 < 0$, the metric is completely smooth. In fact, this geometry has two boundaries, one at $r = 0$ and another at $r = \infty$. Furthermore, the thermal circle takes a minimum value for $g_{uu}$ at $r = \sqrt{-b_0}$. This is precisely the double trumpet with the size of the neck set by $b_0$. One can readily check that $K = 0$ at the neck. Finally, let us make the following important remark. The geometries with $b_0 < 0$ are all diffeomorphic to the $b_0 = -1$ solution as can be seen by rescaling $r$ and $u$ as $r \to \sqrt{-b_0}r$ and $u \to u/\sqrt{-b_0}$. Notice that for $b_0 > 0$ the geometry was already fixed at a particular value for $b_0$ (if fixing the periodicity of $u$) and so we cannot do this rescaling. This is not true for the Lorentzian geometries, where we can do the rescaling for any sign of $b_0$. 

Our primary focus in what follows will be boundaries at large, constant $r$ and we want to fix the extrinsic curvature on those slices.\footnote{In the standard Dirichlet case, this metric can also be used to get the boundary Schwarzian action, see for instance \cite{Gonzalez:2018enk, Godet:2020xpk}} At constant $r = r_0$ slices, we have
\be 
k =  \frac{r_0^2 + b_0}{r_0^2 - b_0}.
\ee
As we will see this greatly simplifies some of the analysis below. An important relation that we will use throughout this paper is that for $b_0>0$ and $r_0 > \sqrt{b_0}$ (or $r_0 < \sqrt{b_0}$) we have that $k>1$, while for $b_0<0$ we find $k<1$.

The geometries at $b_0 > 0$ (and so $b_0 = 1/4$) or $b_0 < 0$ or, equivalently, $k > 1$ or $k <1$, respectively, can also be brought in more conventional form through the following coordinate transformations,
\begin{align}
b_0 &= \frac{1}{4} :\quad r = \frac{1}{2}\coth \frac{\rho}{2} \quad \Rightarrow \quad ds^2  = \frac{d\rho^2 + du^2}{\sinh^2 \rho},\\
b_0 &< 0 : \quad r = \frac{\sqrt{-b_0}}{2} \tan \frac{\rho}{2}, \quad \hat{u} = u\sqrt{-b_0}\quad \Rightarrow \quad ds^2  = \frac{d\rho^2 + d\hat{u}^2}{\sin^2 \rho},
\end{align}
where in the second case $\rho \in [0,\pi)$. In the first case one can do another coordinate transformation to bring the geometry the form,
\be 
ds^2 = (r^2 - 1) d\t^2 + \frac{dr^2}{r^2 - 1}.
\ee
This looks like the familiar Euclidean black hole solution, but we will see below that such an interpretation is misleading.

\begin{figure}
    \centering
    \includegraphics[width=0.8\textwidth]{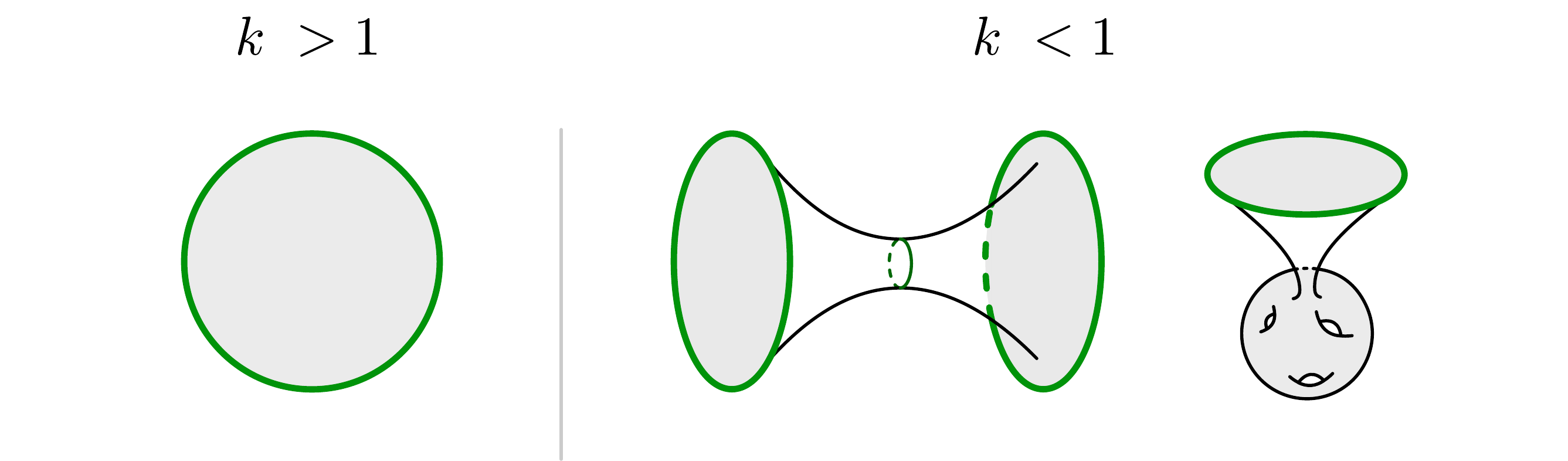}
    \caption{Summary of the classical Euclidean geometries for fixed $K$ and $g_{uu}$ boundary conditions. For $k>1$ we can have a smooth bulk geometry, the disk. For $k<1$ the geometry is cylindrical and we have drawn (in slightly darker green) another fixed $K$ and $g_{uu}$ boundary in the middle and cutting the geometry there gives the trumpet geometry.}
    \label{fig:summaryFixedK}
\end{figure}

Having discussed the solution to the equations of motion, let us now impose the various boundary conditions. Let us first consider the Euclidean theory. We want to fix $K$ and $\int du \sqrt{g_{uu}}$ at some radial slice $r = r_0$:
\be
K|_{r = r_0} = k,\quad \int_{r=r_0} du \sqrt{g_{uu}} = L\,.
\ee

Notice that with these boundary conditions we cannot fix $b_0$ (in case it is negative) to $-1$, since such rescalings would change the boundary length. Solving this for $b_0$ and $r_0$, we find
\be \label{sols}
r_0 = \frac{L(k+1)}{4\pi},\quad b_0 = \frac{L^2 (k^2 -1)}{16\pi^2}\,.
\ee
For $k > 1$ we have $b_0 > 0$ and hence $b_0 = 1/4$, which means that $L$ and $k$ are not independent:
\be 
\label{eq:ND-fixed-L-from-K-disk}
\text{For disk: }\qquad L = \frac{2\pi}{\sqrt{k^2-1}}.
\ee
On the other hand, for $k<1$ only $b_0 < 0$ is a consistent bulk geometry. At this point we can reiterate the important observation that we have made in the previous subsection. We know that for $b_0 > 0$ the geometry has the disk topology, but for $b_0 < 0$ it is cylindrical. So depending on $k > 1$ or $k<1$ we either get a disk or cylinder geometry. In fact, we can also construct the trumpet geometry with the metric \eqref{eq:metricFG} by cutting the geometry at that radial location where the thermal circle becomes geodesic. This happens at $r = \sqrt{-b_0}$. The trumpet geometry is thus \eqref{eq:metricFG} with $r \geq \sqrt{-b_0}$ and conventionally we set the boundary length of the geodesic boundary to $b$,\footnote{This should not be confused with the function $b(u)$ in the Fefferman-Graham metric \eqref{eq:metricFG}.} which we can then relate to $b_0$ as
\be
b = 4\pi \sqrt{-b_0}\,,
\ee
and, hence, for the trumpet geometry we get a relation between $L$, $k$ and $b$:
\be\label{eq:proper-length-fixed-K-trumpet}
\text{For trumpet: }\qquad L = \frac{b}{\sqrt{1-k^2}}.
\ee
Thus, we conclude that only for $k<1$ the trumpet is an allowed (by the boundary conditions) bulk geometry and so only for those values of $k$ higher topologies contribute to the partition function. Such geometries do not contribute when $k>1$. In that case only disks contribute.

Finally, it is also interesting to consider the case when the manifold has a conical defect, which, for convenience, we can place at $r = \sqrt{b_0}$, which is real since $b_0 > 0$ now. In this case, we want $ 4 b_0 = \Theta^2$ where the angular deficit is $2\pi(1 -\Theta)$. In such a case we find, using \eqref{sols}, that
\be 
L = \frac{2\pi |\Theta|}{\sqrt{k^2-1}}\,.
\ee
which is analogous to \eqref{eq:proper-length-fixed-K-trumpet} with the geodesic length $b$ identified with the the deficit angle $\Theta$ -- this makes sense since we can understand this geometry as an analytic continuation of the trumpet geometry.

Before moving on to studying the solution of the dilaton, let us briefly mention the Lorentzian theory. The Lorentzian geometries take the form (again for constant $b(u)$)
\be\label{LorGeom}
ds^2 = \frac{dr^2}{r^2} - \left( r - \frac{b_0}{r}\right)^2 dt^2,
\ee
which, by the rescaling mentioned above, are diffeomorphic to the geometry with $b_0 = 1$ when $b_0 > 0$ and diffeomorphic to the geometry with $b_0 = -1$ whenever $b_0 < 0$. This is possible since the time direction covers the entire real line. Through analytic continuation $u = i t$ (or $\hat{u} = i t$) we can map the Euclidean solutions to Lorentzian ones. For $k > 1$ we get the Lorentzian black hole solution, but as we mentioned before such an interpretation is subtle, whereas for $k < 1$ we get global $AdS_2$. To impose the boundary conditions we fix the boundary metric and not the fixed length (because that is infinite). This is morally the same and gives the same values for $r_0$ and $b_0$ as in \eqref{sols}, but with $L \to 2\pi \a$ if we fix $g_{uu} = \a^2$ at $r = r_0$.

In both the Lorentzian and Euclidean case, the geometry is completely fixed by the boundary conditions. The opposite is true for the dilaton. For these particular boundary conditions, we do not impose any constraint on the dilaton. To gain some intuition of what this means, we can solve the Euclidean bulk equations of motion for the dilaton in the coordinates \eqref{eq:metricFG}. For constant $b(u)$ we have
\be
\label{eq:dilaton}
\phi = \frac{A}{r}\left( 1 + \frac{r^2}{b_0}  \right) + \left(r - \frac{b_0}{r}\right)\left(B \cos(2\sqrt{b_0}u) + C \sin(2\sqrt{b_0} u)\right).
\ee 
where the constants $A$, $B$, and $C$ are arbitrary constants. Since $A$, $B$, and $C$ cannot be fixed from the boundary conditions, we cannot say whether $\phi$ has a minimum, maximum or no local extremum at all. Therefore, since the location at which $\phi$ is at its local minimum is typically interpreted as the location of a 2D black hole horizon, when fixing $K = k$ and $\int du \,\sqrt{g_{uu}} = L$ we cannot specify whether the geometries under consideration contain a black hole or not. We will briefly come back to this  issue in section \ref{sec:discussion}.

Next, we proceed by studying the classical solutions from a different perspective, by reformulating JT gravity as a $PSL(2, \mR)$ gauge theory.  
 
\subsubsection{Connections to the BF formulation of JT gravity and varying $K(u)$}
\label{sec:connection-to-BF}

It is also useful to think about the classical geometries in the gauge theory formulation of JT gravity, in particular when we want to study boundaries with a varying, i.e. time-dependent extrinsic curvature. For the present boundary conditions, there are no boundary terms,\footnote{Apart from counter terms which we can set to zero.} and so the JT action is that of an pure $PSL(2,\mathbb{R})$ BF gauge theory with action \cite{isler1989gauge, chamseddine1989gauge, Blommaert:2018oro, Mertens:2018fds, Saad:2019lba, Iliesiu:2019xuh}
\be
S = -i \int \Tr B F
\ee
with $B$ an adjoint scalar and $F = dA + A \wedge A$ the field strength of an $\mathfrak{sl}(2)$ gauge field. These variables are given in terms of the gravitational variables as (see for instance \cite{Saad:2019lba})
\be 
A =\frac{1}2 \left(\begin{matrix} -e^1 & e^2 - \omega\\ e^2 + \omega & e^1\end{matrix}\right) = \left(\begin{matrix} -\frac{dr}{2r} & r du \\ -\frac{b(u)}{r} du & \frac{dr}{2r} \end{matrix} \right)\,,  \qquad B = -i \begin{pmatrix}
\phi^1 & \phi^2 + \phi \\
\phi^2 - \phi & -\phi^1
\end{pmatrix}
\ee
Here the $e^i$ are the zweibeins of the $2d$ metric and $\w$ the spin connection,
\be 
e^1 = \frac{dr}{r},\quad \quad  e^2 = \left(r - \frac{b(u)}{r}\right) du,\quad \quad \w = - \left(r + \frac{b(u)}{r}\right) du\,.
\ee
The fields $\phi^i$ are additional Lagrange multiplier fields necessary to enforce torsionlessness of the metric. In the second equality we have put in the explicit metric einbeins in the Fefferman-Graham gauge \eqref{eq:metricFG}. The equations of motion for $B$ give us flatness of the $\mathfrak{sl}(2,\mathbb{R})$ connection and is the analogue of the $R+2 = 0$ equation. There is a beautiful way to solve such flatness conditions: $A = g^{-1}dg$. Let us directly study the general case of time-dependent $b(u)$ and consider,
\be 
g = \begin{pmatrix}
r^{-1/2}\psi_1'(u) & r^{1/2}\psi_1(u) \\
r^{-1/2}\psi_2'(u) & r^{1/2}\psi_2(u) 
\end{pmatrix}.
\ee
The flatness of the connection then tells us that 
\be 
\psi_{1,2}''(u) + b(u) \psi_{1,2}(u) = 0.
\ee
This differential equation is known as Hill's equation and is a central object in the study of Schrodinger operators with a periodic potential (recall $b(u)$ is periodic) and coadjoint orbit theory \cite{Balog:1997zz}. Next, we need to make sure that $g$ is an $PSL(2,\mR)$ element, which is nothing but a normalisation of the Wronskian,
\be
\det g = 1 \quad \Leftrightarrow \quad \psi_1' \psi_2 - \psi_1 \psi_2' = 1.
\ee
Thus $\psi_{1,2}$ need to be linearly independent solution to the Hill equation such that the Wronskian is normalised to unity. Finally, to ensure that the connection corresponds to a smooth bulk geometry, meaning that for the disk topology the boundary is contractible, the holonomy of the gauge field $A$ needs to be trivial. We define the holonomy $U$ to be
\be 
U = \mathcal{P} \exp\left(-\oint A \right) = g^{-1}(2\pi)g(0),
\ee
where in the second equality we used flatness of $A= g^{-1} dg $. Triviality of $U$ thus means $g(0) = \pm g(2\pi)$ (recall we are working with $PSL(2,\mathbb{R})$ here). Other geometries with a different topology can be obtained by considering a different value for the holonomy. 

Let us exemplify this method of finding smooth geometries by taking constant $b$. It is straightforward to check that the holonomy is given by (at some radius $r$)
\be \label{eq:constantbholonomy}
U = \begin{pmatrix}
\cos(2\pi \sqrt{b_0}) & -\frac{r}{\sqrt{b_0}}\sin(2\pi \sqrt{b_0})\\
\frac{\sqrt{b_0}}{r} \sin(2\pi \sqrt{b_0}) & \cos(2\pi \sqrt{b_0})\\
\end{pmatrix}.
\ee
So, when $\sqrt{b_0} = n/2$ for $n\in \mathbb{N}$, $U = \pm \mathbf{1}$ and we have as smooth disk geometry. Furthermore, we know that $\sqrt{b_0}$ is related to $k$ and $L$ as given in \eqref{sols} and so one can easily see that this integer $n$ corresponds to a geometry with a 'boundary' that winds $n$ times. This is not a smooth orientable geometry and is therefore not present in the gravity path integral and as a result only $n = 0, 1$ are relevant here. In fact, for $n=0$ we are dealing with the Poincar\'{e} patch, which has a real line as its boundary and so clearly does not have the topology of a disk. This case is therefore excluded as well. The only relevant geometry is thus $n = 1$. 

We can also extract the other geometries we found before. For the three different geometries, disk, defect and wormhole, we found the three different conjugacy classes $U$ can be in. To determine them, we simply look at the trace of $U$:
\be 
\Tr\,U = 2 \cos( 2\pi \sqrt{b_0}).
\ee
The conjugacy classes are then determined as follows,
\be 
\Tr\,U = \left\{\begin{array}{lll} 
< 2 & \quad b_0 > 0, b_0 \neq n^2/4 & \text{Elliptic} \\
= 2 & \quad b_0 = n^2/4 & \text{Parabolic}\\
> 2 & \quad b_0 < 0 & \text{Hyperbolic}
\end{array} \right..
\ee
An elliptic $U$ thus corresponds to the defect geometry, a parabolic one to the disk and the wormhole is realised by an holonomy in the hyperbolic conjugacy class. This concludes our discussion of the constant $b(u) = b_0$ solutions. For varying $b(u)$ it is much more complicated to find solutions, in particular due to the path ordering involved in the definition of $U$. However, for the disk topologies, we need to find periodic/anti-periodic solutions to Hill's equatin and we can ask about the existence of such solutions.

For a generic potential $b(u) = r^2 \tilde{K}(u)$, the Hill equation does not have such solutions. However, so far, we have not fixed the value of $r$ at which the boundary is located; therefore, we wish to understand whether there exist values of $r$ such that the Hill equation $ \psi''_{1,2}  + r^2 \tilde K(u)\, \psi_{1, 2}  = 0 $ has periodic/anti-periodic solutions. In turn, if there exists a value of $r$ for which such a solution exists, then this in turn fixes the values of the proper length $L$. Following the intuition developed for the case with constant $K=k $ where only configurations with $k>1$ exist on the disk, we will first focus on the case $k(u)>1$ and consequently, $\tilde K(u) >0$. For such a case, assuming that $\tilde K(u)$ is sufficiently differentiable, Lyapunov \cite{Lyapunov} proved that there exist an infinite series of values of $r$ for which the Hill equation has a periodic/anti-periodic solution. The proof relies on the Liouville transformation; i.e.~solutions of  $ \psi''_{1,2}  + r^2 \tilde K(u)\, \psi_{1, 2}  = 0 $ can be mapped to solutions of 
\be
\label{eq:Hill-to-spectral-eq}
\frac{d^2\xi_{1,2}}{dx^2} + (r^2 \gamma^2 + Q(x)) \xi_{1,2} = 0\,,
\ee
via the map
\be
x = \frac{1}\gamma \int_0^u \tilde K^{1/2}(\tilde u) d\tilde u\,, \qquad \gamma = \frac{1}\pi \int_0^{2\pi}   K^{1/2}(\tilde u) d\tilde u\,,\qquad \xi(x) = \tilde K^{\frac{1}4}(u) \psi(u)\,,
\ee
and
\be 
Q(x) = (\tilde K(u))^{-\frac{1}4}  \frac{d^2(\tilde K(u))^{\frac{1}4} }{dx^2}\,.
\ee
Eq.~\eqref{eq:Hill-to-spectral-eq} can be viewed as an eigenvalue equation for the differential operator $\frac{d^2}{dx^2} + Q(x)$ with eigenvalue $-r^2 \gamma^2$. Since, when $Q(x)$ and $\gamma$ are real, such an operator has an infinite number of real eigenvalue when imposing periodic or anti-periodic solutions and it then follows that there are an infinite number of values of $r$ that yield a manifold which is smooth. As can be seen from the case of constant $k$, different values of $r$ correspond to  different windings of the boundary. Since we are interested in surfaces that have a boundary which is not self-intersecting we will solely be interested in a single value (out of the infinite series) of $r$, which we call $r_s$. Consequently, the proper length is fixed to 
\be
\label{eq:proper-length-function-of-tildeK}
L = r_s \int_0^{2\pi} du\, |1-\tilde K(u)|\,.
\ee

One can similarly prove that when $0<k(u) <1$ ($\tilde K(u)<0$) then the Hill equation does not have any periodic or anti-periodic solutions. Therefore, in that case the boundary conditions exclude all surfaces with the topology of a disk. 

To summarize, while we have not found the analytic constraint that relates $k(u)$ to $L$ (since we have not explicitly determined $r_s$ in \eqref{eq:proper-length-function-of-tildeK})  we have proven that given $k(u)>1$, there always exists a single value of $L$ for which a hyperbolic surface with disk topology exists. For $-1<k(u)<1$ the disk topology never exists, therefore generalizing the results found for constant $k$. 

\subsection{Quantum theory}

\subsubsection{The disk topology}

Let us now compute the partition function for fixed $K =k$ and $g_{uu}$ boundary conditions. We have seen that with these boundary conditions no boundary terms are necessary and when setting the counter terms $\mC_i$ to zero, we just need to path integrate over the bulk action. We thus wish to calculate (in Euclidean signature),
\be 
Z_{ND}[k,\ell] = \int \frac{\mathcal{D}\phi\mathcal{D}g}{\Vol({\rm diff})}\,e^{\phi_0 \chi(\mathcal{M})} e^{\int_{\mathcal{M}} d^2 x \sqrt{g}\,\phi(R+2)}. 
\ee
The manifolds we sum over in this path integral are denoted by $\mathcal{M}$. For now will restrict our attention to orientable manifolds with a single boundary and discuss multiple boundaries below. Here the contour of $\phi$ is chosen to be along the imaginary axis, in which case the path integral over $\phi$ gives a delta functional $\d(R+2)$ and our path integral simply counts the number of hyperbolic metrics having a boundary with fixed $k$ and $\ell$ modulo diffeomorphisms:
\be 
Z_{ND}[k,\ell] = \int \frac{\mathcal{D}g}{\Vol({\rm diff})}\, e^{\phi_0 \chi(\mathcal{M})} \;\delta(R+2).
\ee
The easiest way to compute this partition function is by going to a BF theory formulation of JT gravity that reviewed above.\footnote{In \cite{Saad:2019lba} this was done for closed manifolds, which then also has no boundary terms, but the boundary conditions are different, so we cannot directly apply their formulae to the current case.} The path integral is then just a function of the holonomy $U$ of the gauge field $A$ around the boundary circle,
The partition function of the $PSL(2,\mathbb{R})$ gauge theory is \cite{Saad:2019lba}
\be 
Z_{ND}[U] = \int \frac{\mathcal{D}A\mathcal{D}B}{\Vol({\rm gauge})}\;e^{\phi_0 \chi(\mathcal{M})}e^{i \int \Tr BF} = \int \frac{\mathcal{D}A}{\Vol({\rm gauge})} e^{\phi_0 \chi(\mathcal{M})}\delta(F),
\ee
with $\chi(\mathcal{M} = D_2) = 1$. This partition function thus computes the number of flat connections modulo gauge transformations, but with one important caveat. In the gauge theory only connections $A$ that are not related by smooth gauge transformations are counted seperately, whereas in the gauge theory variables such connections might be related by large diffeomorphisms, i.e. elements on the mapping class group of the underlying manifold. In the gravity theory metrics related by the mapping class group are not counted separately and in the gauge theory we need to account for that. This imposes no subtleties in the disk for which the mapping class group is trivial, but does so for the cylinder or higher topologies as the mapping class group is non-trivial in those cases. The disk partition function $Z_{\rm Disk}[k,L]$ for constant $k$ ($b(u) = b_0$) can thus be computed rather directly in the $PSL(2,\mathbb{R})$ gauge theory. In fact, there is only one flat $PSL(2,\mathbb{R})$ connection (modulo gauge transformations) on the disk and the partition function is
\be 
\label{eq:delta-function-BF}
Z^{\rm (Disk)}_{ND}[U] = e^{\phi_0}\delta(U-\mathbf{1}).
\ee
To write this in terms of the diffeomorphism invariant gravity variables $k$ and $\ell$, we need to express the holonomy $U$ in terms of $k$ and $\ell$. The holonomy was given in \eqref{eq:constantbholonomy} and so the remaining step is to rewrite the $\delta$-function on the $PSL(2, \mR)$ group manifold in terms of Dirac-delta functions solely dependent on $L$ and $k$. There are two possible ways in which one can rewrite the $\delta$-function on the group manifold, dependent on the space of test-functions which we plan to integrate the test function against.   The first, (a) are general functions on the group manifold, while the second (b) are trace-class functions. If we fix $PSL(2, \mR)$ gauge transformations on the boundary (or, equivalently, fixing gravitational diffeomorphisms), then, in order to get some generic observable, one can integrate the partition function against a general function on the group manifold, while if one does not fix gauge transformations on the boundary, the holonomy is not gauge invariant but its conjugacy class is. To summarize, the interpretation of the $\delta$-function on the group as a distribution in $L$ and $K$ depends on whether we fix diffeomorphisms on the boundary.

In appendix \ref{sec:delta-functions-app} we give a detailed account about the difference between the resulting conversion for the $\delta$-function on the group manifold. When studying the distribution (a) we find that \eqref{eq:delta-function-BF} becomes
\be 
\label{eq:Disk-ND-part-function-dist-a}
Z^{\rm (Disk)}_{ND}[k,\ell] = \frac{e^{\phi_0}}{2\pi} \delta^2(0) \delta \left(\sqrt{b_0} - \frac{1}2 \right) = \frac{2e^{\phi_0}}{\sqrt{k^2-1}} \delta^2(0)\delta\left(L - \frac{2\pi}{\sqrt{k^2-1}}\right),
\ee
where we used \eqref{sols}.  When studying the distribution (b) that acts on the space of trace-class functions, we instead find
\be 
\label{eq:Disk-ND-part-function-dist-b}
Z^{\rm (Disk)}_{ND}[k,\ell] = \frac{e^{\phi_0}}{2\pi}\delta''\left(\sqrt{b_0} - \frac{1}{2}\right) = \frac{e^{\phi_0}}{2\pi} \delta''\left(\frac{L\sqrt{k^2-1}}{4\pi} - \frac{1}{2}\right)\,.
\ee
We can perform a further check of this result by going in between Dirichlet-Dirichlet and Dirichlet-Neumann boundary conditions using a Laplace transform on the boundary value of the dilaton field $\phi_r(u)$. In the limit in which the proper length of the boundary $L$ is large (i.e. the Schwarzian limit when studying Dirichlet-Dirichlet boundary conditions) the path integral that we need to evaluate is (omitting the topological term)
\be 
\label{eq:ZND-from-Schw}
Z_{ND}^\text{(Disk)} &= \int D\phi_b(u) e^{-2 \int_0^\beta du \sqrt{g_{uu}}\phi_b(u) k} \sim \int  D\phi_r(u) \int D f(u) e^{\int_0^\b du \phi_r(u) (\Sch(f, u) - \kappa)}\nn \\  &\sim 
\int D f(u) \delta (\Sch(f,u)-\kappa) 
\ee
where the "$\sim$" means equality up to counterterms which have the role to eliminate an overall divergent constant multiplying the partition function. Furthermore, above, we define the renormalized quantities  $k = 1 + \varepsilon^2 \kappa$ and $\phi_r(u) = \phi_b(u)/\varepsilon$.  
In appendix \ref{app:SchwarzianFixedK} we compute the path integral \eqref{eq:ZND-from-Schw} explicitly and show that the result agrees with the BF computation from \eqref{eq:Disk-ND-part-function-dist-a}.

This concludes our brief discussion of the disk partition function with fixed $k$ and $L$ boundary conditions. For varying $K_{r=r_s} = k(u)$ we can use the results about the Hill equation with general potential we discussed above, in particular \eqref{eq:proper-length-function-of-tildeK} and we get
a $\delta$-function that activates when $L - \int_0^{2\pi} du\, \left|r_s -\frac{k(u)}{r_s}\right| = 0$. As previously mentioned, for a given $k(u)$, we cannot determine $r_s$ analytically; rather as described in section \ref{sec:connection-to-BF}, it is determined by the solution of the Hill equation.

\subsubsection{The trumpet and cylinder with fixed $K$}

We can compute the partition function on the trumpet or cylinder, again, in two different ways.  

In the first, we again rely on the BF formulation of JT gravity and need to glue the opposite sides of the gravitational theory when placed on a manifold with the topology of a square. This is very similar to the computation of the partition function of the conventional BF theory on a cylinder with one important distinction.  This  is that in the gravitational theory we should also quotient by the mapping class group of the cylinder (given by $\mathbb Z$) while in conventional BF theory this is not necessary. As we will see, considering this quotient is important in obtaining a convergent partition function. 

In the second approach, we will again transform the partition function with DD boundary conditions to that with ND boundary conditions via a Laplace transform. This amount to evaluating the Schwarzian path integral \eqref{eq:ZND-from-Schw} on a different orbit than we did in the subsection above; i.e. $\text{Diff}(S^1)/U(1)$ instead of $\text{Diff}(S^1)/SL(2, \mathbb R)$. 

We begin by presenting the first approach. Formally we want to compute  
\be
\label{eq:gluing-for-JT-cylinder}
Z_{ND}^{\rm (Cylinder)}(U_L, U_R) =  \int_{PSL(2, \mR)/\mathbb Z} dh  \,\delta(U_L h^{-1} U_R^{-1} h)\,, 
\ee
where $U_L$ and $U_R$ are the holonomies along the edges of the cylinder and $h$ is the holonomy along a line uniting the two sides. 
The difficulty in computing this partition function is partly due to the quotient and in particular in explaining what quotient of $PSL(2,\mR)$ we need to consider. To figure this out, we will consider a particular on-shell metric solution, convert it to the gauge theory variables, and compute the holonomies around the non-contractible cycle and along the gluing curve. From the metric variables, it will be clear what identifications we need to make in order to quotient out the mapping class group and the above procedure will tell us what part of the gauge theory variables, in particular the group element, needs to be identified.

To that end, it is first useful to consider an on-shell hyperbolic configuration which has the following metric \cite{Saad:2019lba} 
\be 
\label{eq:Dehn-twist-metric}
ds^2 = d\rho^2 + \cosh(\rho)^2 [b dx + \tau \delta(\rho) d \rho]^2\,, \qquad x\sim x+1\,.
\ee
The variable $\tau$ represents the distance of a twist made along the closed geodesic at $\rho = 0$. To see that the metric \eqref{eq:Dehn-twist-metric} is equivalent to a purely hyperbolic metric one can introduce the coordinate $y$:
\be 
y=  bx + \tau \theta(\rho)\,, \qquad dy = b dx + \tau \delta(\rho) d \rho \,.
\ee
In such a case, it is clear that $\tau$ should be identified up $\tau \sim \tau+b$ since the coordinate $x$ is compact and periodically identified. In fact, shifts of $\tau$ by $b$ are precisely identified with the Dehn twists which generate the mapping class group $\mathbb Z$ of the cylinder. Therefore, in our computation of the cylinder partition function we will need to identify holonomies for which $\tau$ is shifted by $b$.

To achieve this we start by writing the frame and spin-connection for the metric \eqref{eq:Dehn-twist-metric}, 
\be 
\begin{split}
    e^1 = d \rho\,&, \qquad  e^2 = b \cosh(\rho) dx + \tau \cosh(\rho) \delta(\rho) d \rho,\qquad \omega = - \sinh(\rho)(b dx+\tau \delta(\rho) d\rho), \\[0.25cm]
    &\qquad\qquad A={1\over 2}\left(\begin{matrix}
    -d\rho & b e^{\rho} dx+\tau e^{\rho} \delta(\rho)d\rho\\
    b e^{-\rho}dx+\tau e^{-\rho}\delta(\rho)d\rho & d\rho
    \end{matrix}\right).
\end{split}
\ee
and consider the holonomy around the closed cycle of the cylinder at the left and right boundary, 
\be 
\label{eq:holonomy-cylinder}
U_{L,R} = \cP \exp\left(-\oint_{\rho = \text{const}} A\right) = \left(\begin{matrix}
\cosh(b/2) & -e^{\rho_{\text{bdy}_{L, R}}} \sinh(b/2) \\ 
-e^{-\rho_{\text{bdy}_{L, R}}} \sinh(b/2) & \cosh(b/2)
\end{matrix}\right)
\ee
whose eigenvalues are $( e^{b/2}, \,e^{-b/2})$ for all $\rho$. Thus, we see that the conjugacy class of the holonomy around the closed cycle of the cylinder is independent of the location at which the holonomy is evaluated. This of course follows from the fact that the connection is flat and from the gluing formula \eqref{eq:gluing-for-JT-cylinder} we clearly see that the $\delta$-function only activates when $U_L$ and $U_R$ are in the same conjugacy class, regardless of the integration space for $h$. To understand the quotienting procedure we consider the Wilson line along a curve with constant $x$, from one end of the cylinder to the other. In such a case, 
\be 
\label{eq:gluing-holonomy}
h = \cP \exp\left(-\int_{\rho= \rho_{\text{bdy}_L}}^{\rho = \rho_{\text{bdy}_R} } A \right)  = \left(\begin{matrix}
e^{\frac{\rho_{\text{bdy}_R} - \rho_{\text{bdy}_L}}2}\cosh(\tau/2) & - e^{\frac{\rho_{\text{bdy}_R} + \rho_{\text{bdy}_L}}2}\sinh(\tau/2) \\ - e^{-\frac{\rho_{\text{bdy}_R} + \rho_{\text{bdy}_L}}2}\sinh(\tau/2) &  e^{\frac{\rho_{\text{bdy}_L} - \rho_{\text{bdy}_R}}2}\cosh(\tau/2)
\end{matrix}
\right)
\ee
where we assume $\rho_{\text{bdy}_L} < 0$ and $\rho_{\text{bdy}_R}>0$.\footnote{This assumption is unimportant since one can always perform a diffeomorphism which places the $\delta$-function in \eqref{eq:Dehn-twist-metric} at any $\rho$-location. } We will thus assume that the holonomy along the two-sides of the cylinder take the form \eqref{eq:holonomy-cylinder} where $b$ is determined by the proper length and extrinsic curvature on each side while the ``gluing'' holonomy is given by \eqref{eq:gluing-holonomy} up to a $\mathbb Z$ identification. It is convenient to re-express \eqref{eq:gluing-for-JT-cylinder} in terms of the conjugacy classes of $U_L$ and $U_R$. To diagonalize $U_L$ and $U_R$ we have that 
\be 
Z_{ND}^{\rm (Cylinder)}(U_L, U_R)  = \int dh \delta( \tilde U_L A_L h^{-1} A_R^{-1}\tilde U_R^{-1} A_R^{-1} h A_L^{-1}),
\ee
where $U_{L,R}  = A_{L, R}^{-1} \tilde U_{L, R} A_{L, R}$
\be
A_{L, R} = \frac{1}{\sqrt{2}}\left( \begin{matrix}
-e^{-\rho_{L, R}/2} & e^{\rho_{L, R}/2}\\
-e^{-\rho_{L, R}/2} & -e^{\rho_{L, R}/2} 
\end{matrix}  \right),
\ee
and where we can change integration variables to $\tilde h = A_L h^{-1} A_R^{-1}$,
\be\label{eq:JTcylinder}
Z_{ND}^{\rm (Cylinder)}(U_L, U_R)  = \int d\tilde{h} \delta( \tilde U_L\tilde h \tilde U_R \tilde h^{-1})\,.
\ee
For the on-shell configuration studied above  $\tilde h$ can be expressed as 
\be 
\label{eq:decomposing-h}
\tilde h  = e^{-\tau \s_3/2}\,, 
\ee
and thus (since we are studying an on-shell configuration) for these values of $\tilde U_L$, $\tilde U_R$ and $\tilde h$ the $\delta$-function activates. 

Nevertheless, this computation allows us to see which elements $h$ or $\tilde h$ are related by a large diffeomorphism.  Namely, we have learned that the quotient we want to do is on elements in $PSL(2,\mathbb{R})$ which are related by multiplication by a hyperbolic diagonal group element. To be more specific, given two elements $h_1$ and $h_2$, we want to identify them when they are related by a right multiplication by the element $e^{-b\s_3/2}$. Concretely, let us consider the $KNA$ decomposition of a general element of $PSL(2,\mathbb{R})$,
\be 
\label{eq:Iwasawa-decomp}
\tilde h =  K N A\,, \qquad K = \left(\begin{matrix}
\cos(\theta) & \sin(\theta)\\-\sin(\theta)&\cos(\theta)
\end{matrix}\right)\,, \qquad N = \left(\begin{matrix}
1 & n \\ 0 & 1
\end{matrix} \right)\,, \qquad A = \left(\begin{matrix}e^{a} & 0\\ 0 & e^{-a} \end{matrix}\right)\,,
\ee
for which the Haar measure can be written as $d\tilde h  = d\theta \, dn \,  da $. Thus our identification is a restriction on the range of $a$, which we take to be from $0$ to $b/2$ (while in $PSL(2, \mR)$ this range is non-compact).

Let us now return to the general problem, i.e. evaluating \eqref{eq:JTcylinder}. Using the above decomposition the delta function of a group element in the $KNA$ decomposition is given by $\delta(U) = \delta(\theta) \delta (n) \delta(a)$. We take $U = \tilde U_L\tilde h \tilde U_R \tilde h^{-1}$ with $\tilde{h}$ written in the $KNA$ decomposition and consider (the possibly off-shell configuration) $\tilde U_{L, R} = \exp(\sigma_3 \l_{L, R}/2)$. The only thing left to do is to evaluate the one-loop factor, which can be done straightforwardly by expanding the argument of the delta function close to $\theta = 0$ and $n = 0$, since the delta function only fires when $\tilde{h}$ is diagonal
\be  
\delta(\tilde U_L \tilde h \tilde U_R \tilde h^{-1} ) = \delta \left(\begin{matrix}
e^{(\l_1-\l_2)/2} &  2 e^{\l_1/2}(n+\theta)\sinh(\l_2/2) \\ 
2 e^{-\l_1/2} \theta \sinh(\l_2/2) &e^{-(\l_1-\l_2)/2}
\end{matrix}\right) + \mathcal{O}\left(n \theta,\, n^2, \,\theta^2\right)
\ee
The coordinate $a$ does not appear in the integrand and directly gives a factor of $b/2 = \l_L/2 = \l_R/2$. The cylinder partition function is thus, 
\be\label{eq:cylinder}
Z_{ ND}^{\rm (Cylinder)}(\l_L, \l_R) = \l_L \frac{\d(\l_L - \l_R)}{4 \sinh^2 \l_L/2}\,,
\ee
where $e^{\pm \l_L/2}$ and $e^{\pm \l_R/2}$ are the eigenvalues of the holonomies $\tilde{g}_L$ and $\tilde{g}_R$.

Naively, the trumpet partition function would be defined by taking $\l_L = b$, i.e. the geodesic boundary has a holonomy determined by $b$. However, this is too quick, because the cylinder is obtained by gluing two trumpets and there can be a non-trivial measure factor appearing in the gluing depending on how we define the trumpet. In fact, since here we have expressed the partition function as a delta function on the conjugacy classes of the boundary holonomies, there is indeed a non-trivial measure as can be seen from the Weyl integration formula for $PSL(2,\mathbb{R})$, see appendix \ref{sec:delta-functions-app}. Furthermore, if we would have constructed two trumpets from \eqref{eq:cylinder}, we would have implicitly taken two twist integrals (the integrals over $a$ that gave us a factor of $b/2$) into account, whereas we need only one. To account for that, we divide by $b/2$ in the gluing of two trumpets. We thus get
\be
Z^{\rm (Cylinder)}_{ND}(\l_L,\l_R) = \int_0^{\infty} db \left(2 \sinh^2 b/2\right) \left(b \frac{\d(b - \l_L)}{4 \sinh^2 \l_L/2}\right)\left( b \frac{\d(b - \l_R)}{4 \sinh^2 \l_R/2}\right)\left(\frac{2}{b}\right),
\ee
where each term in brackets corresponds to the measure obtained from the Weyl integration formula,\footnote{Here we picked a particular normalisation $(\alpha=1)$ of the measure instead of keeping it around as was done in \cite{Saad:2019lba}.} the two cylinders, and the division by $b$ to account for overcounting, respectively. From this, we define the trumpet partition function by distributing the measure (the first term in round brackets) in the definition of the trumpet,
\be 
\label{eq:ND-trumpet}
Z_{ND}^{\rm (Trumpet)}(b,\l_L) = \frac{\d(b - \l_L)}{2\sinh b/2},
\ee
and the gluing measure is just the Weil-Petersson measure $b\,db$.

In appendix \ref{app:SchwarzianFixedK} we confirm this way of defining the trumpet and cylinder partition function by doing the path integral directly using the boundary Schwarzian mode. The main difference between the computation in for the trumpet as opposed to the disk \eqref{eq:ZND-from-Schw} is the number of zero modes for the Schwarzian field $F(u)$ (since we are integrating over different orbits). Because of this difference the divergent factor $\delta^2(0)$ present on the disk in \eqref{eq:Disk-ND-part-function-dist-a} is no longer present in the case of the trumpet or cylinder.

To complete our analysis we solely need to use the relation between the conjugacy class of the holonomy for the Neumann-Dirichlet boundary and the extrinsic curvature $K$ together with the proper length $L$. We will take the $\l_L$ to be the eigenvalue of this holonomy while $\l_R = b$ is the eigenvalue of the holonomy on the closed geodesic boundary. This is given by 
\be 
\label{eq:ND-trumpet-part-function}
Z^{(\text{Trumpet})}_{ND}(b;L,k) = \frac{\delta\left( L \sqrt{1-k^2} - b\right)}{2 \sinh b/2}.
\ee
The cylinder partition function then becomes,
\be 
\label{eq:ND-cylinder-part-function}
Z^\text{(Cylinder)}_{ND}(L_1, k_1;L_2, k_2)  = \frac{L_1\sqrt{1-k_1^2}}{4\sinh^2\left(\frac{L_1 \sqrt{1-k_1^2}}2\right)} \delta\left(L_1\sqrt{1-k_1^2} - L_2 \sqrt{1-k_2^2}\right).
\ee 

\subsubsection{Putting it all together}
\label{sec:ND-quant-putting-it-all-together}

We can now consider the general genus expansion of the partition function of JT gravity with ND b.c. We have already seen that boundaries which have $K> 1$ lead to factorization: 
\be 
\label{eq:part-function-factorization}
\<Z_{ND}&(k_1>1, L_1) \dots Z_{ND}(k_i>1, L_i]) Z_{ND}(k_{i+1}<1, L_{i+1})\dots Z_{ND}(k_n<1, L_n)\> =  \\ &= \<Z_{ND}(k_1>1, L_1)\> \dots \<Z_{ND}(k_i>1, L_i)\>  \<Z_{ND}(k_{i+1}<1, L_{i+1})\dots Z_{ND}(k_n<1, L_n)\>\nn
\ee
with 
\be
\label{eq:ND-disk-part-function-summary}
\<Z_{ND}&(K_1>1, L_1)\> =  \frac{e^{\phi_0}}{2\pi} \delta''\left(\frac{L\sqrt{k^2-1}}{4\pi} - \frac{1}{2}\right)
\ee
Thus, we simply want to discuss the genus expansion of $\<Z_{ND}(K_{1}<1, L_{i+1})\dots Z_{ND}(K_n<1, L_n)\>$. As previously specified, we simply have to integrate the partition function the trumpet against the Weil-Petersson volumes (denoted here by $\Vol_{g,n}(b_1, \dots, b_n)$ for a manifold of genus $g$ and with $n$ geodesic boundaries with lengths $b_1$, $\dots$ $b_n$) using the Weil-Petersson measure. This yields, 
\be\label{eq:pertZKguucorr}
\<Z_{ND}&(K_{1}<1, L_{1})\dots Z_{ND}(K_n<1, L_n)\> \sim \nn \\ &\sim  \sum_{g=0}^{\infty} e^{\phi_0 \chi_{g, n}}\int db_1 b_1 \cdots \int db_n b_n \,\frac{\delta( L_1 \sqrt{1-K_1^2} - b_1 )}{2 \sinh b_1/2} \cdots \frac{\delta( L_n \sqrt{1-K_n^2} - b_n)}{2 \sinh b_n/2}\nn \\ &\qquad \qquad \qquad\qquad \qquad \qquad \times \Vol_{g, n}(b_1, \dots, b_n) \nn \\& = \frac{\left( {L_1 \sqrt{1-K_1^2}} \,\cdots \,  {L_n \sqrt{1-K_n^2}}\right)}{2^n \sinh \frac{L_1 \sqrt{1-K_1^2}}2 \,\cdots \, \sinh \frac{L_n \sqrt{1-K_n^2}}2} \sum_{g=0}^{\infty} e^{\phi_0 \chi_{g,n}}\Vol_{g, n}(L_1 \sqrt{1-K_1^2}, \dots, L_n \sqrt{1-K_n^2})
\ee
where we formally define $\Vol_{0,1}(b) \equiv 0$ and $\Vol_{0,2}(b_1, b_2)\equiv  \delta(b_1-b_2)/b_1$.\footnote{This corresponds to the Laplace transforms of the volumes $W_{g, n}(z_1, \dots, z_n)$ to have $W_{0,1}(z) = 0$ and $W_{0, 2}(z_1, z_2) = \frac{1}{(z_1-z_2)^2}$.} Thus, up to overall constants the partition function for boundaries with $K_i < 1$ simply yields a sum over the Weil-Petersson volumes. We will use the simplicity of this result to determine the matrix integral interpretation for the insertion of an ND boundary in the JT gravity path integral.  

\subsection{Matrix integral interpretation}

\label{sec:ND-matrix-integral-interp}

With the geometric result for the partition function in mind, we now wish to find the operator insertion in the JT gravity  matrix integral which reproduce the results above. To do this, we will first attempt to understand this matrix integral interpretation by performing a path integral for the boundary valued dilaton $\phi_b(u)$ for the  matrix integral result with Dirichlet-Dirichlet boundary conditions. While we shall not be able to perform this path integral exactly, we will still be able to determine an integral kernel which when applied to the Dirichlet-Dirichlet partition function reproduces the factorization properties observed in the previous subsection. Using this kernel, we will be able to finally determine the correct operator insertion in the matrix integral to reproduce \eqref{eq:part-function-factorization}--\eqref{eq:pertZKguucorr}.  

\subsubsection{A Transformation Kernel}

As previously mentioned, $\phi_b(u)$ and $K(u)$ are canonical conjugates. Thus, to obtain $Z_{ND}$ from $Z_{DD}$, we need to perform the path integral
\be 
Z_{ND}[k, L] =  \int D\phi_r(u) e^{\frac{1}{\e^2}\int_0^\beta  du \phi_r(u)(1-k(u))} Z_{DD}[\phi_b(u), L].
\ee
To continue, we will use the formula for  $Z_{DD}[\phi_b(u), L]$ for varying $\phi_b(u) = \phi_r(u)/\varepsilon$, obtained in the Schwarzian nearly-AdS$_2$ limit ($L = \b/\varepsilon \to \oo$) by interpreting the results in appendix C of \cite{Stanford:2019vob} or the results reviewed in more detail in appendix A in \cite{Iliesiu:2020zld}: \footnote{In this subsection we fix the normalization of the matrix integral over $H$ following the convention of \cite{Saad:2019lba} with $\gamma  = 1$.}
\be
Z_{DD}^{\rm (Schw)}[\phi_b(u), L] = e^{\int_0^\beta {du}\frac{\phi_r'(u) ^2 }{2\phi_r(u)}} \left\<\Tr \, e^{-  H \int_0^\beta  du/\phi_r(u)}\right\>_\text{MI}\,,
\ee
where $\left\<\Tr e^{-  H \int_0^L du/\phi_r(u)}\right\>_\text{MI}$ is the expectation value of the ``partition function operator'' with the effective temperature $\beta_\text{eff}[\phi_r(u)]  = \int_0^\b du/\phi_r(u)$. Thus, we would like to evaluate 
\be 
\label{eq:integral-transform-ND-to-DD}
Z_{ND}[k(u), L] = \int d\tilde L &\left\<e^{-  H \tilde L}\right\>_\text{MI} \nn \\ &\times  \underbrace{\int_{-i\infty}^{i \infty}  d\sigma \int D\phi_r(u) e^{ \sigma\left(\tilde L - \int_0^\beta \frac{du}{\phi_r(u)} \right)}e^{\int_0^\beta {du}\frac{\phi_r'(u) ^2 }{2\phi_r(u)}}  e^{\frac{1}{\e^2}\int_0^{\beta} du \phi_r(u)(1-k )}}_{ \cF_k(\tilde L,L)}\,,
\ee
where we have introduced the Lagrange multiplier $\sigma$ to emphasize that one can obtain the ND partition function by performing an integral over $L$ (instead of a path integral over $\phi_r(u)$) with the appropriate kernel $\cF_k(\tilde L, L)$. This allows us to give an exact definition of the kernel in the Schwarzian lmit. However, the path integral over $\phi_b(u)$ in the last line of \eqref{eq:integral-transform-ND-to-DD} is difficult to perform exactly. Nevertheless, a naive semi-classical evaluation of this path integral in the Schwarzian limit yields (integrating solely over configurations for which $\phi_r(u)$ varies only very slowly with $u$)
\be 
\label{eq:approx-kernel}
 \cF_k^\text{semi-classical}(\tilde L,L) \sim \frac{1}{ \tilde L^{3/2}} e^{L^2\frac{(1-k)}{\tilde L}}\,.
\ee
It is tempting to guess that the factorisation properties required of the kernel can be reproduced by a simple modification of the approximate semi-classical kernel \eqref{eq:approx-kernel}. In what follows, we will show that this is the case. Moreover, we will see that this kernel can also be applied away from the Schwarzian regime ($k\approx 1$, $L\to\infty$).

The modification of the kernel which we propose is:
\be \label{eq:nd-kernel}
\mathcal{F}_K(\tilde L ,L) =   \sqrt{2\pi} \frac{e^{\frac{\lambda^2}{2\tilde L} }}{\tilde L^{3/2}}\mu(\lambda) \qquad \qquad\mu(\lambda) = \left( \frac{\lambda}{2\sinh\frac{\lambda}{2}} \right)
\ee 
where we have defined 
\be 
\lambda \equiv L\sqrt{1-k^2} \,.
\ee
In the Schwarzian limit, $k=1+O(\varepsilon^2)$ we may also replace $1-k^2 \sim 2(1-k)$ in this expression to recover the approximate kernel \eqref{eq:approx-kernel}. 

From the BF perspective, $\lambda$ is the eigenvalue of the holonomy along the corresponding boundary. When $k<1$, $\lambda$ is real, otherwise it is purely imaginary.  
We will demonstrate that this kernel satisfies the desired properties. This allows us to define the ND partition function as a simple integral transform of the DD partition function,
\be \label{eq:nd-convolution}
    Z_{ND}(k,L) = \int\limits_{\mathcal{C}} d\tilde L \,\, Z^{(g)}_{DD}(\tilde L) \mathcal{F}_K(\tilde L,L).
    \ee
 where $\mathcal{C}$ is a contour to be specified. Here, it is useful to think of the perturbative (in $e^{-\phi_0}$) part of the decomposition of the Dirichlet partition function into contributions from different genera 
    \be 
    \label{eq:ZDD-expansion}
    \langle Z_{DD}(\tilde L) \rangle = e^{\phi_0} Z_{DD}^{\rm (Disk)}(\tilde L) + \sum\limits_{g=1}^{\infty}e^{(1-2g)\phi_0} \int\limits_0^{\infty}db\, b\, V_{g,1}(b) Z_{DD}^{\rm (Trumpet)}(b;\,\tilde L) 
    \ee
  where we have the explicit expressions for the disk and trumpet \cite{Saad:2019lba, Stanford:2019vob} 
        \be 
Z_{DD}^{\rm (Disk)}(\tilde L) =  \frac{1}{\sqrt{2\pi}}\frac{e^{\frac{2\pi^2}{\tilde L}}}{ \tilde L^{3/2}} \qquad\qquad Z_{DD}^{\rm (Trumpet)}(b;\,\tilde L) = \frac{1}{\sqrt{2\pi}} \frac{e^{-\frac{b^2}{2\tilde L }}}{\tilde{L}^{1/2}}.
\ee
where we are studying the theory with the effective temperature, $\tilde L = \beta/\phi_r$, and where $b$ is the proper length of the closed geodesic that separates the trumpet from the rest of the bordered higher genus Riemann surface.   
    
    In order to study the properties of \eqref{eq:nd-convolution}, let us first consider the trumpet contribution. In order to specify the contour of integration for the kernel, it is useful to introduce the variable $z=\tilde{L}^{-1}$. We then chose the integration contour for $z$ along the imaginary axis
       \be \label{eq:dd-higher-genus}
 \frac{ Z_{ND}^{\rm (Trumpet)}(k,L)}{\mu(\lambda)} =&  \int\limits_{-i\infty}^{i\infty} \frac{dz}{z^2}  \,\,  \left( z^{1/2} e^{-\frac{b^2}{2 }z } \right)
    \,  \left( z^{3/2} e^{\frac{\lambda^2}{2 }z }\right)\nonumber = 
    \int\limits_{-i\infty}^{i\infty} dz \, e^{\frac{z}{2}\left(-b^2+L^2 (1-k^2) \right)} \\
    =&\left\{\begin{array}{ll}
    2 \delta\left( b^2 - \lambda^2 \right) = \frac{1}{\lambda} \delta\left( b-\lambda\right) & \text{for }k < 1 \\ 
    0 &\text{for }k > 1
    \end{array}\right..
    \ee 
  When $k\geq 1$, the integral contour for $z$ can be closed on the right half plane $\text{Re}(z)>0$ and hence vanishes. Thus, the trumpet does not contribute for an ND boundary with $k>1$. In the opposite case, $k<1$, we have the second line which reproduces \eqref{eq:ND-trumpet} derived from other methods. Thus, the trumpet result  
    \be 
Z_{ND}^{\rm (Trumpet)}(k,L) = \frac{\delta(b-L \sqrt{1-k^2})}{2\sinh \frac{b}2},
    \ee
agrees with \eqref{eq:ND-trumpet} obtained from BF theory or by considering the Laplace transform of the Dirichlet-Dirichlet partition function. Therefore, applying the kernel \eqref{eq:nd-convolution} to \eqref{eq:dd-higher-genus} we obtain the sum over Weil-Peterson volumes from \eqref{eq:pertZKguucorr}. 

    The final step needed to compare the matrix integral and  geometric results is to understand the effect of the kernel on manifolds with disk topology. The disk contribution to \eqref{eq:nd-convolution} is given by
    \bea \label{eq:nd-disk}
    \frac{ Z_{ND}^{\rm (Disk)}(k,L)}{\mu(\lambda)} =  \int\limits_{-i\infty}^{i\infty} \frac{dz}{z^2} \,\,  \left( z^{3/2} e^{2\pi^2 z} \right)
    \,  \left(z^{3/2} e^{\frac{\lambda^2}{2} z } \right) &=& 
    \int\limits_{-i\infty}^{i\infty} dz \,z\,  e^{z\left(2\pi^2+\frac{L^2}{2} (1-k^2) \right)}
    \eea
     In this case, when $k\leq 1$, the integral contour for $z$ can be closed on the left half plane $\text{Re}(z)<0$ and hence vanishes. Thus the disk topology does not contribute to the partition sum in this case. When $k>1$, the integral can be evaluated exactly and yields the same type of divergence as in \eqref{eq:ND-disk-part-function-summary}. Therefore, we find that the integral kernel \eqref{eq:nd-kernel} reproduces the ND partition function computed in section \ref{sec:ND-quant-putting-it-all-together}.

\subsubsection{Operator Insertion in Matrix Integral}
Given that we see that the kernel applied to the DD partition functions reproduces the correct ND partition function we can now ask what operator in the matrix  integral directly reproduces the insertion of this latter boundary in the gravitational path integral. This is simply given by considering the action of the kernel on the partition function operator:
\be 
\label{eq:ND-matrix-integral-repr}
e^{-\tilde{L} H} \to \int_{-i\infty}^{+i \infty} \frac{dz}{z^{1/2}}\,\mu(\lambda)\,e^{L^2\frac{(1-K^2)}{2}z} &e^{- \frac{H}{z}} \sim
\begin{cases}
\frac{\cos ( L\sqrt{1-K^2} \sqrt{2H})}{\sinh\left(\frac{L}{2}\sqrt{1-K^2}\right)} & K<1, \\
 \frac{\cosh ( L\sqrt{K^2-1} \sqrt{2H})}{\sin\left(\frac{L}{2}\sqrt{K^2-1}\right)} & K>1
\end{cases}\,.
\ee
Thus, we conclude that 
\be \label{eq:ND-matrix-operator}
Z_{ND}[K, L] \sim \<
\Tr \,\cos (\lambda \sqrt{2H})\>_{MI},
\ee
which has the same properties as the ones discussed above. 

It is important to note however, that one needs to keep careful track of the integration contour when evaluating this operator via analytic continuation in the matrix integral. For instance, when integrating the $K<1$ operator against the trumpet, the ``energy of the Hamiltonian'' $H$ should be integrated along the imaginary axis.

\section{Fixing $K$ and $\partial_n \phi - \phi K$: a less rigid geometry }
\label{sec:NN-bc}

The study the partition function with fixed $K \equiv k(u)$ and $\partial_n \phi - \phi K \equiv - \phi_b'(u)/2$ (NN boundary conditions)\footnote{The function $\phi_b'(u)$ should not be confused with the derivative of the boundary value of the dilaton which is not specified for these boundary conditions.} is similar, yet less rigid than that for fixed $K$ and $g_{uu}$. To emphasize this point we start by re-analyzing the classical behavior discussed in the previous section. While fixing $K$ completely fixes the geometry of the manifold, by setting the location and proper length of the boundary, when fixing $g_{uu}$ the arbitrary constants in the classical dilaton solution \eqref{eq:dilaton}  could not be fixed from the ND boundary condition. For the NN boundary condition however, fixing $\partial_n \phi - \phi K$ fixes the constant $A$ in  \eqref{eq:dilaton}, however the constants $B$ and $C$ are still unfixed in the classical solution.  Thus, $B$ and $C$ are zero modes for the dilaton solution.  In such a case, the classical dilaton solution in Euclidean or Lorentzian signature \eqref{eq:dilaton} is set (in the Fefferman-Graham gauge \eqref{eq:metricFG}) to be:
\be 
r_0 &=\sqrt{b_0 \frac{k+1}{k-1}}\,,\nn\\
\phi(r,u) &=  \frac{\phi_b'}{2r\sqrt{k^2-1}}\left( 1 + \frac{r^2}{b_0}  \right) + \left(r - \frac{b_0}{r}\right)\left(B \cos(2\sqrt{b_0}u) + C \sin(2\sqrt{b_0} u)\right)\,,
\ee 
with the boundary located at $r_0$.
Nevertheless, as for the ND b.c.~, for $k>1$ we again only have the disk contribution, and $b_0= 1/4$ in order for the Euclidean geometry to be smooth. For $k<1$ and, consequently, $b_0 < 0$, we only have contributions from higher genus or multi-boundary geometries.

Finally, in order to compare the classical results to the partition function which we shall obtain shortly, it is useful to compute the on-shell action (for disk topologies) coming from boundary term in \eqref{eq:NN-action}:
\be
\label{eq:NN-on-shell}
S_{NN}^\text{on-shell} = -L \phi_b' = -\frac{2\pi}{\sqrt{k^2-1}} \phi_b'\,.
\ee
where we have used the fact that fixing $K = k$ also fixes the proper length of the boundary $L$, according to \eqref{eq:ND-fixed-L-from-K-disk}.

We now discuss the quantization of the gravitational theory with the NN boundary conditions. We have seen in the previous section that fixing $k>1$ completely fixes the topology of the manifold and the proper length of its boundary. Therefore, since the proper length of the boundary is not fixed by the boundary conditions, the partition function on the disk will no longer yield a $\delta$-function even though the geometry is fixed. Similarly, for $k<1$ we again only receive contributions from higher genus or multi-boundary manifolds. Thus, we again have that 
\be 
\label{eq:NN-part-function-factorization}
&\<Z_{NN}(k_1>1, \phi'_{b,1}) \dots Z_{NN}(k_i>1,\phi'_{b,i}) Z_{NN}(k_{i+1}<1, \phi'_{b,i+1})\dots Z_{NN}(k_n<1, \phi'_{b,n})\>\nn =  \\ &= \<Z_{NN}(k_1>1, \phi'_{b,1})\> \dots \<Z_{NN}(k_i>1, \phi'_{b,i})\>  \<Z_{NN}(k_{i+1}<1,\phi'_{b,i+1})\dots Z_{NN}(k_n<1, \phi'_{b,n})\>.
\ee
However, while in the previous section, fixing $g_{uu}$ on the boundary completely fixes the length $b$ of the closed geodesic homotopic to the boundary, in this section, since the proper length of the boundary is not fixed, $b$ is also, consequently, not fixed.

To make things concrete and obtain the partition function, $Z_{NN}$, we can go between Dirichlet-Dirichlet and Neumann-Dirichlet boundary conditions using a Laplace transform and integrating over the boundary metric. Schematically this is given by\footnote{Below, for the NN boundary conditions  we are no longer fixing $g_{uu}$. Therefore, we need to fix $k(u)$ in a diffeomorphism invariant fashion. This can be done by going to a diffeomorphism gauge where $g_{uu}$ is constant and where the periodicity of $u$ is fixed. Thus, when one fixes $k(u)$ for the ND and NN boundary conditions in \eqref{eq:ND-to-NN-transform}, it should be understood that $u$ is specified in this gauge. }
\be 
\label{eq:ND-to-NN-transform}
Z_{NN}[k(u), \phi_b'] = \int \frac{\mathcal{D}g_{uu}}{\text{Diff}(S^1)} e^{2\int du \sqrt{g_{uu}}(\phi K - \partial_n \phi) } Z_{ND}\left[k(u), \int du \sqrt{g_{uu}}\right] .
\ee
The ND system has boundary diffeomorphism symmetry which has been mode out to get a finite result.
Just like in the case of path integral of a relativistic particle, 
the integration over the boundary metric can be simplified by going to the gauge where $\sqrt{g_{uu}}$ is equal to a constant.
Then the functional integral of $\sqrt{g_{uu}}$ can be reduced to an integral over the proper length $L$.\footnote{See section 9.2 of \cite{Polyakov:1987ez}  for discussion about the procedure of fixing boundary diffeomorphism symmetry.} With the existence of a marked (for instance, at $u=0$) point at the boundary, we have\footnote{If we don't have a marked point, there will be an additional ${1\over \beta}$ factor coming from the time translation symmetry. } 
\be 
\label{eq:ND-to-NN-transform2}
Z_{NN}[k(u), \phi_b'] = \int dL e^{2\int du \sqrt{g_{uu}}(\phi K - \partial_n \phi) } Z_{ND}[k(u), L]  .
\ee
For simplicity, we will assume that $k(u) = k$ and $\phi_b'(u)=\phi_b'$ are constant. In such a case, we find that for the disk: 
\be 
Z_{NN}^{\text{(Disk)}}(k>1, \phi_b') &= \int dL e^{  L \phi_b'} Z_{ND}^{\text{(Disk)}}(k>1, L) \nn 
\\ &= \frac{e^{\phi_0}}{2\pi} \int dL e^{L \phi_b'} \delta''\left(\frac{L\sqrt{k^2-1}}{4\pi} - \frac{1}2\right) \,,
\ee
where we have used \eqref{eq:Disk-ND-part-function-dist-b} for the ND disk partition function, acting on the space of trace-class functions.  Consequently, 
\be
\< Z_{NN}(k>1, \phi_b')\>&= \frac{8\pi e^{\phi_0} e^{\frac{2\pi \phi_b'}{\sqrt{k^2-1}}}}{(k^2-1)^2}\,.
\ee
This matches with the classical saddle-point obtained in \eqref{eq:NN-on-shell}. 
Similarly, for the trumpet we find 
\be 
Z_{NN}^{\text{(Trumpet)}}(k<1,\, \phi_b')&=  \int dL  e^{L \phi_b'} \frac{\delta(L \sqrt{1-k^2  }- b)}{2\sinh \frac{b}2}\nn \\  &= \frac{e^{\frac{b\phi_b'}{\sqrt{1-k^2}}}}{2 \sqrt{1-k^2} \sinh \frac{b}2}\,,
\ee
where we have used \eqref{eq:ND-trumpet-part-function} for the ND partition function of the trumpet. Therefore, the NN partition function when summing over connected manifolds with $n$ boundaries is 
\be 
\label{eq:NN-connected-JT-part-function-k<1}
\<&Z_{NN}(k_{1}<1, \phi_{b, 1}')\dots Z_{NN}(k_n<1, \phi_{b,n}')\>_\text{conn.}  \sim \nn \\ 
&\sim \sum_{g \geq 0}e^{\phi_0 \chi_{g, n}}\int db_1\,b_1 \int db_2 \, b_2 \dots \int db_n b_n \frac{e^{-\sum_{i=1}^n \frac{b_i \phi_{b, i}'}{\sqrt{1-k_i^2}} }}{2^n \left(\prod_{i=1}^n  \sqrt{1-k_i^2} \sinh \frac{b_i}2 \right)} \Vol_{g,n} (b_1, \dots, b_n).
\ee
Note that with the conventions from section \ref{sec:ND-quant-putting-it-all-together} for the cylinder we have that $\Vol_{0,2} (b_1, b_2) = \delta(b_1-b_2)/b_1$. Therefore, the NN cylinder partition function exhibits a log-divergence coming from the $\sim 1/b$ behavior in the limit $b\to0$ of the integrand. Therefore, if we do not consider a more general UV completion of JT gravity,\footnote{It would however be interesting to understand whether there indeed is a UV completion of the model which leads to a convergent cylindrical partition function for the NN boundary conditions. 
} all partition functions with $n \geq 2$ boundaries are dominated by cylindrical contributions. 

Next, with the results in section \ref{sec:ND-matrix-integral-interp}, we can determine what operator insertion in the matrix integral description of JT gravity yields the NN results for the partition function obtained in \eqref{eq:NN-part-function-factorization}--\eqref{eq:NN-connected-JT-part-function-k<1}. By considering the Laplace transform of \ref{eq:ND-matrix-integral-repr} we have that
\be 
\label{eq:NN-operator-interp}
Z_{NN}(k, \phi_b') &\leftrightarrow \int dL e^{-L \phi_b'}\, \Tr\, \frac{\cos\left(L \sqrt{1-k^2} \sqrt{2 H}\right)}{\sinh\left(\frac{L}2 \sqrt{1-k^2}\right)} \nn \\ &= \Tr\left[ - \frac{\mathcal{H}\left(-\frac{1}{2} +\frac{\phi_b'}{\sqrt{1-k^2}}- i\sqrt{2H}\right)+\mathcal{H}\left(-\frac{1}{2} +\frac{\phi_b'}{\sqrt{1-k^2}}+ i\sqrt{2H}\right)}{\sqrt{1-k^2}}\,\right]\,,
\ee
where $\mathcal{H}(x)$ is the analytically continued harmonic number.

\section{Fixing $\phi$ and $\partial_n \phi$: microcanonical ensemble, relation with eigenbranes}

\label{sec:DN-bc-eigenbranes}
In this section, we will discuss the DN boundary condition in JT gravity: fixed boundary value of dilaton field $\phi(u)|_{\partial \cM} = \phi_b$ and its normal derivative $\partial_n\phi|_{\partial \cM} = \phi_b'$. Let's start by  discussing the classical solution with the DN boundary condition.
Again, using the Fefferman-Graham gauge, we have the bulk solution of the dilaton field \eqref{eq:dilaton} on a disk:
\be
\phi = \frac{A}{r}\left( 1 + 4r^2  \right) + \left(r - \frac{1}{4r}\right)\left(B \cos(u) + C \sin( u)\right).
\ee
Since we are fixing the boundary value of dilaton and its derivative, we can perform an $SL(2,\mathbb{R})$ coordinate transformation to set $B=C=0$.
Then the constant $\phi_b$ and $\phi_b'$ conditions fix $A$ and put the boundary at constant $r=r_0$. Solving for these two conditions gives,
%\JK{is $r_0$ large here? Maybe we want to scale $\phi_b$ and $\phi_b'$ in a particular way? Do we want to write $\phi_b'$ here? }
\be
A={\sqrt{\phi_b^2-\phi_b'^2}\over 4},\qquad~~r_0={1\over 2}\sqrt{\phi_b+\phi'_b\over \phi_b-\phi_b'}.
\ee
Notice that $\phi_b $ and $\phi_b'$ scale like $1/\varepsilon$ and their difference as $\varepsilon$.

The quantization of the theory is closely related to that with the standard DD boundary condition where the boundary value of dilaton $\phi_b$ and the boundary metric $g$ are fixed.
As we have explain previously, in Euclidean signature, the DD boundary condition with constant $\phi_b$ corresponds to the canonical ensemble of the underlying gravitational theory where the inverse temperature $\beta$ is the total regularized length $\varepsilon \int du \sqrt{g_{uu}}$.

The DN boundary condition, can be obtained from a Laplace transform of the DD boundary conditions. In this case we have to perform an integral of the DD partition function over the boundary metric $g_{uu}$ with an additional weighting $e^{\varepsilon E \int du \sqrt{g_{uu}}}$:
\be
Z_{DN}(\phi_b, E)=\int {\frac{\mathcal{D} g_{uu}}{ \text{Diff}(S^1)}} e^{\varepsilon E\int du \sqrt{g_{uu}} } Z_{DD}(\phi_b, \,\varepsilon\int \sqrt{g_{uu}}).
\ee
Due to a counter term in the DD partition function, the relation between $E$ and $\partial_n\phi$ is 
\be
E={2(\phi_b-\phi_b')\over \varepsilon}.
\ee
We will again follow the gauge fixing procedure for the boundary metric described in section \ref{sec:NN-bc} to write\footnote{Here the $\b$ integration contour is along the imaginary axis.}
\be
Z_{DN}(\phi_b, E)=\int d\beta e^{E\beta} Z_{DD}(\phi_b, \beta).
\ee
Given that in the matrix integral description we know that considering a DD boundary corresponds to inserting a partition function operator in the matrix integral, $ Z_{DD}(\phi_b, \beta) \leftrightarrow \Tr e^{-\beta H}$, we then can use this Laplace transform to obtain the matrix integral description of an ND boundary. This corresponds to the insertion of the density of states operator in the matrix integral,\footnote{The  normalization of $H$ in the matrix integral on the RHS of \eqref{eq:DN-to-eigenbranes} is the same as in \cite{Saad:2019lba} when setting $\gamma=\phi_r$. }
\be
\label{eq:DN-to-eigenbranes}
Z_{DN}(\phi_b, E) \qquad \leftrightarrow \qquad 
\rho(E)=\Tr \,\delta (H-E)=\sum_{i}\delta(\lambda_i-E).
\ee
It is interesting to make connection of the DN boundary with the energy-eigenbrane dicussed in \cite{Blommaert:2019wfy}, which fixes one eigenvalue of the matrix $H$ to be some fixed value $E$: $\delta(\lambda_1-E)$.
Using the permutation symmetry of the eigenvalues in the random matrix distribution, we see that the the DN partition function is just proportional to the eigenbrane operator. Consequently, any correlator measured in JT gravity with  DN boundary conditions will be equivalent to computing the same correlator in the matrix integral in the presence of eigenbranes. For this reason, we can identify the DN boundary conditions as energy-eigenbranes or energy-branes for short.

We now turn to the exact computation of the partition function for the DN boundary conditions. As in \eqref{eq:ZDD-expansion}, the DD partition function can be written as a summation of Riemann surfaces \cite{Saad:2019lba}:
\bea \label{eqn:ZDD}
\langle Z_{DD}(\phi_b, \beta)\rangle= e^{\phi_0}Z_{DD}^{\text{(Disk)}}(\phi_b, \beta)+\sum_{g=1}^{\infty}e^{\phi_0(1-2g)}\int_0^{\infty} b db\, V_{g,1}(b) Z_{DD}^{\text{(Trumpet)}}(b;\phi_b, \beta)\,.
\eea
To obtain the microcanonical (ND) partition function, one only needs to inverse laplace transform the corresponding disk and trumpet partition function:
\be
Z_{DN}^{\text{(Disk)}}(E)= {\phi_r\over 2\pi^2}\sinh{2\pi\sqrt{2\phi_r E}};~~~~Z_{DN}^{\text{(Trumpet)}}(E,b)={\phi_r^{1/2}\cos{b \sqrt{2\phi_r E}}\over \pi\sqrt{2E}}.
\ee
The moduli integral and $V_{g,1}(b)$ will not change, since those are fully determined by the bulk curvature constraint and the Weil-Petersson measure.
Replace these expressions in \ref{eqn:ZDD}, we have the full genus expansion of the DN partition function,
\bea \label{eqn:ZDN}
\langle Z_{DN}(E)\rangle= e^{\phi_0}Z_{DN}^{\text{(Disk)}}(E)+\sum_{g=1}^{\infty}e^{\phi_0(1-2g)}\int_0^{\infty} b db \, V_{g,1}(b) Z_{DN}^{\text{(Trumpet)}}(E,b)
\eea
Since we take the asymptotic limit $\phi_b\rightarrow \infty$, the integral range of $b$ is from zero to infinity.
Finally, note that non-perturbative corrections for correlators of the form \eqref{eqn:ZDN} were computed in the context of energy-branes in \cite{Blommaert:2019wfy}. With these results in mind we proceed to discuss the implications of these results to black hole toy models and, in particular, their relation to the microcanonical thermofield double and to fixed area states.

\section{Discussion, applications, and future directions}

\label{sec:discussion}
 
We can now use the results derived in sections \ref{sec:Kguu}--\ref{sec:DN-bc-eigenbranes} in the setting of black hole thermodynamics, holography, cosmology, of the minimal string, and of the baby universe Hilbert space where JT gravity has served as a toy model in the past \cite{Almheiri:2019qdq, Penington:2019kki,Maldacena:2019cbz, Cotler:2019nbi, Chen:2020tes}.

 In the case of black hole thermodynamics, we will only focus on the case of fixed $\phi$ and $\partial_n \phi$ in which a black hole horizon is present. We again emphasize that for fixed $K = k$, one cannot know whether the geometry has a horizon or not since the solution for $\phi$ cannot be fully fixed. Consequently, the factorization properties (that we point out in the previous sections) are not in tension with the Page curve analysis from  \cite{Almheiri:2019qdq, Penington:2019kki} for which the contribution of connected replica wormholes is required. 
 
 Nevertheless, as we will see, the fixed $K = k$ and $g_{uu}$ boundary condition is natural in the context of cosmological toy models, where similar factorization properties can be derived. For the convenience of the reader, it is useful to rephrase these factorization properties (both in AdS$_2$ and dS$_2$) in the baby universe Hilbert space language of Marolf and Maxfield \cite{Marolf:2020xie}; in turn, this will motivate the renaming of the fixed $K$ boundary conditions as $\alpha$-eigenbranes, or $\alpha$-branes for short. 
 
 Furthermore, it will prove informative to interpret the change of boundary conditions that we have studied throughout this paper in the larger context of AdS/CFT. In that case, understanding the change of boundary conditions in the bulk amounts to studying the flow of the boundary theory under a multi-trace deformation. While for JT gravity, the exact boundary dual is still uncertain, it is interesting to speculate about the role of such deformations in the SYK model. 
 
 Finally, we will comment on the relation with $(2,p)$ minimal string theories. Specifically,  in \cite{Saad:2019lba}, it was shown how the spectral density of JT gravity arises from the large $p$ limit of the $(2,p)$ minimal string. Here, we show how JT gravity's different boundary conditions can be explicitly mapped to boundary conditions in the minimal string. 

\subsection{Microcanonical Thermofield Double State and Fixed Area State}

\label{sec:micro-thermofield}

In this section, we will review the microcanonical thermofield double (MCTFD) state proposed by Marolf \cite{Marolf:2018ldl}, which is closely related with our DN boundary condition, and discuss its connection with the fixed area state \cite{Dong:2018seb, Akers:2018fow}.
We will argue that, in some instances, the MCTFD serves as a better boundary definition of the Fixed Area State.

Let's imagine embedding JT gravity into a UV complete theory where the spectrum is discrete (either by considering an SYK realization or by embedding the 2D theory into a higher dimensional UV complete theory of gravity). 
If we consider the Lorentzian continuation of the path integral with DD boundary condition, the Euclidean part of the path integral prepares the thermofield double state (TFD) of a two-sided black hole.  Similarly, the DN boundary condition admits an analogous state called the microcanonical thermofield double state \cite{Marolf:2018ldl}
\be
|E\rangle=\sum_{i} f(E-E_i)|i\rangle|i\rangle,
\ee
where $f(E-E_i)$ is a peaked distribution that approaches a delta function with bandwidth $\Delta E$.
The MCTFD state admits a semi-classical description of a black hole as long as the energy window contains  order $e^{\sim \phi_0}$ of nearby black hole micro-states. Up to a normalization factor, the MCTFD state can be understood as a superposition of thermofield double states evolved with different Lorentzian times, which we denote by $\ket{\beta + i t}$, 
\be
|E\rangle=\int dt e^{i E t} \tilde f(t)|\beta+it\rangle;~~~~~\tilde f(t)=\int dE e^{-it E}f(E).
\ee
The fourier transformation of $f(E)$ effectively constrains the integral range of $t$ to be order ${1\over \Delta E}$ by the uncertainty relation.
If we approximate the distribution $f(E)$ as a square function of width $\Delta E$, then it's clear that the MCTFD has a flat entanglement spectrum with the entanglement entropy between two sides given by
\be
S=S(E)+\log \Delta E,
\ee
where $S(E)$ is the thermal entropy of the black hole.
For other types of smeared distributions of $f$, this qualitative feature will not change.
Notice that as long as $\Delta E$ is order one, the entanglement entropy is close to the thermal entropy. This, of course, is consistent with the semi-classical description.

In the literature, another type of gravitational state with flat entanglement spectrum was also proposed: the fixed area state \cite{Dong:2018seb, Akers:2018fow}.
The fixed area state is a state prepared by Euclidean gravitational path integral such that the area operator on a certain Ryu-Takayanagi surface $\gamma$ is fixed.
Although no explicit prescription on the boundary was given in the past, the authors of \cite{Dong:2018seb}
argued that the fixed area states should have a flat entanglement spectrum by considering the gravitational path integral of the $n$ replica geometry
\be
Z_n=\int \mathcal{D}g_n|_{A_{\gamma}(g_n)=\hat A}\,\,e^{-I(g_n)}\,.
\ee
For the fixed area state, the area of $\gamma$ in the $n$ replica geometry is kept to be $\hat A$, and this means that the classical saddle of the $n$ replica geometry will be just the $n$ copies of the original geometry $g_1^c$ glued at $\gamma$.
This means that the partition function $Z_n$ has saddle point approximation
\be
Z_n=e^{-n I_\text{away}(g_1^c)+(1-n){\hat A\over 4 G}}.
\ee
Here $n I_\text{away}(g_1^c)$ is the contribution away from $\gamma$ and $(1-n){\hat A\over 4 G}$ is the contribution from $\gamma$ with conical angle $2\pi n$.
This shows that the leading (large $N$) Rényi entropy is a constant, and therefore the entanglement spectrum is flat
\be
S_n = {\hat A\over 4 G}\,.
\ee
If we have matter, then there will be order one corrections.

Let's now consider the fixed area version of the TFD state and choose the extremal surface to be the black hole horizon.
By the following relation between the area operator and the modular Hamiltonian \cite{Faulkner:2013ana, Jafferis:2014lza, Jafferis:2015del}
\be
S=\langle K\rangle= \langle A\rangle+S_\text{matter},
\ee
we can think of fixing the area as fixing the eigenvalue of the modular operator $K$ (which in this case is the boundary Hamiltonian). Up to some order one fluctuations, this is the same as the microcanonical thermofield double state.
This leads us to two closely related semiclassical states: the fixed area state and the microcanonical thermofield double state.
Both states describe the same classical geometry and have the same flat entanglement spectrum, but their gravitational prescriptions are different. 

In the description of the fixed area state, one needs to specify the bulk value of the extremal area (in this case the horizon value of the dilaton field) by hand, while for the MCTFD one only specifies the boundary condition, which is a more conventional holographic prescription.
\begin{figure}
    \centering
    \includegraphics[width=0.5\textwidth]{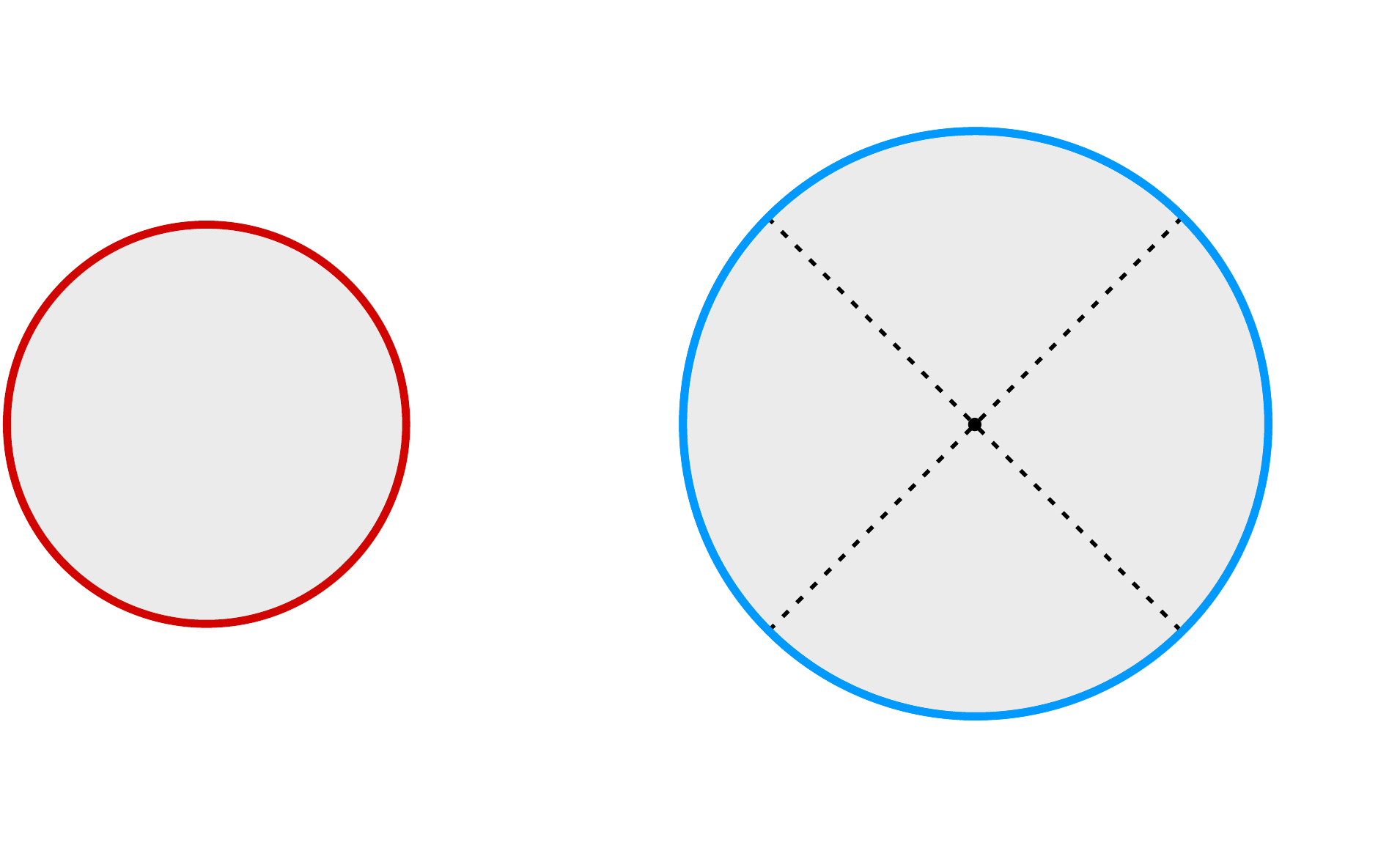}
    \caption{Left: replica geometry for the microcanonical thermofield double state. Right: replica geometry for the fixed area state.}
    \label{fig:nreplica}
\end{figure}
For instance, the $n$ replica geometry of the microcanonical thermofield double state is given by saddle point of the DN partition function which is independent of $n$ and is everywhere smooth
\be
Z_{n}(\text{MCTFD})=\int_{-i T}^{i T} \prod_{i=1}^nd\beta_i  e^{E\sum_{i=1}^n\beta_i}\,\,Z_{DD}\left(\sum_{i=1}^n\beta_i\right)=\left({1\over \Delta E}\right)^{n-1} Z_{DN}(E)\,.
\ee
where $\Delta E$ is again the inverse of the integral range (T) of $\beta$ .
On the other hand, the fixed area state prescription for $n\geq 1$, see figure \ref{fig:nreplica}, has a conical angle at the horizon, very much like the tensor network models discussed in \cite{Pastawski:2015qua, Hayden:2016cfa}.
One can imagine extending the definition of the MCTFD to more general cases where the extremal surface is not the horizon by replacing the boundary Hamiltonian with the corresponding modular Hamiltonian. In such a case, a similar discrepancy will appear.
This leads to a puzzle that apparently the same boundary states can have different (and maybe equivalent) bulk descriptions.
We will leave this to future work.

\subsection{Cosmology}
\label{sec:cosmology}

So far, we have discussed mostly the anti-de Sitter case in two dimensions, but the attentive reader might have noticed that our analysis in section \ref{sec:class-bdy-cond} applies to general dilaton potentials. In particular, it also holds for potentials with positive cosmological constant, where for instance, we have the action 
\be \label{dSAction}
S_\text{top}+S_\text{bulk} = -\chi(\mathcal{M})\phi_0 + \int_{\mathcal{M}} \sqrt{g} \phi (R - 2) - 2\int_{\partial\mathcal{M}} \sqrt{\g} \phi K.
\ee
This theory was considered in detail in \cite{Maldacena:2019cbz}. Instead of time-like asymptotic boundaries as in the AdS case, we now have space-like asymptotic boundaries, and we can ask what type of boundary conditions we can put on the phase space variables there since such variables are identical to those in AdS$_2$. While the variables are the same, there are a few differences. First of all, the object we compute in de Sitter is not a partition function but rather a wavefunctional. Second, in AdS, there are time-like boundaries, and so we can interpret some of the boundary conditions as different ensembles in quantum mechanics, but for dS it is not clear how to do that. For instance, fixing the normal derivative of the dilaton in the AdS case corresponded to fixing the energy of the boundary quantum mechanics, but in dS such an identification is not possible. From the gravitational path integral point of view, such boundary conditions are however well-defined. 

One very natural and common boundary condition in cosmology (in 2D) fixes the metric and extrinsic curvature on the Cauchy slice. In particular, the extrinsic curvature can be used as a time coordinate conjugate to the volume of the spatial slice \cite{PhysRevLett.28.1082}. For JT in dS$_2$, these boundary conditions mean that we fix the extrinsic curvature and metric at the asymptotic past and future. This boundary condition was already discussed extensively in section \ref{sec:Kguu}, so we will be brief here and highlight the important implications for dS$_2$. 

If we solve the equations of motion coming from varying \eqref{dSAction}, we find the following global metric for dS$_2$ and the following profile for the dilaton:
\be\label{globaldS}
ds^2 = -d\t^2 + \cosh^2 \t\,d\varphi^2 ,\quad \phi = \phi_h \sinh \t,
\ee
where $\varphi \sim \varphi + 2\pi$ for the Hartle-Hawking state. In the Fefferman-Graham gauge used in equation \ref{eq:metricFG} we can write the metric solution as
\be \label{FGdS}
ds^2 = -\frac{dt^2}{t^2} + \left(t + \frac{f(\tilde \varphi)}{t}\right)^2 d\tilde \varphi^2.
\ee
Here $t$ and $\tau$ are to be thought of as complex, representing a Euclidean section when purely imaginary and Lorentzian when real and a particular contour in the complex $t$ or $\tau$ plane corresponds to a particular gluing of a Lorentzian section to a Euclidean one. For instance, when computing the Hartle-Hawking wavefunction we employ the no-boundary proposal and continue the metric \eqref{globaldS} (at $\tau = 0$ we set $\tau = i\theta$) to a hemisphere to cap off the space-time.  

As noted in \cite{Maldacena:2019cbz}, since $R = 2$ for these metrics, it seems, naively that we cannot have higher genus contributions. This can be overcome by taking a different contour for the metric. There the contour $\tau = i \pi /2 + \tilde{\t}$ was used \cite{Maldacena:2002vr} and sends the metric in \eqref{globaldS} to 
\be 
ds^2 = -(d\tilde{\t}^2 + \sinh^2 \tilde{\t} d\varphi^2),
\ee
which was dubbed $-\text{AdS}$ for obvious reasons. This then naturally paves the way for considering geometries like the trumpet but with an overall sign in front of the metric. In this way, one can indeed have higher genus contributions to the wavefunction. From the Fefferman-Graham gauge, equation \eqref{FGdS}, it is also amusing to see that the usual contour would already work. If we take $t = i\tilde{\tau}$, then the metric becomes
\be 
ds^2 = - \left( \frac{d\tilde{\tau}^2}{\tilde{\t}^2} + \left(\tilde{\t} - \frac{f(\tilde\varphi)}{\tilde{\tau}}\right)^2 d\tilde \varphi^2 \right).
\ee
This is precisely minus the metric \eqref{eq:metricFG}. So the minus trumpet would be obtained by taking $f(\tilde\varphi)$ constant and negative and cut the geometry at the minimal waist size. There we can then glue higher genus geometries. 

\begin{figure}[t]
    \centering
    \includegraphics[width=0.8\textwidth]{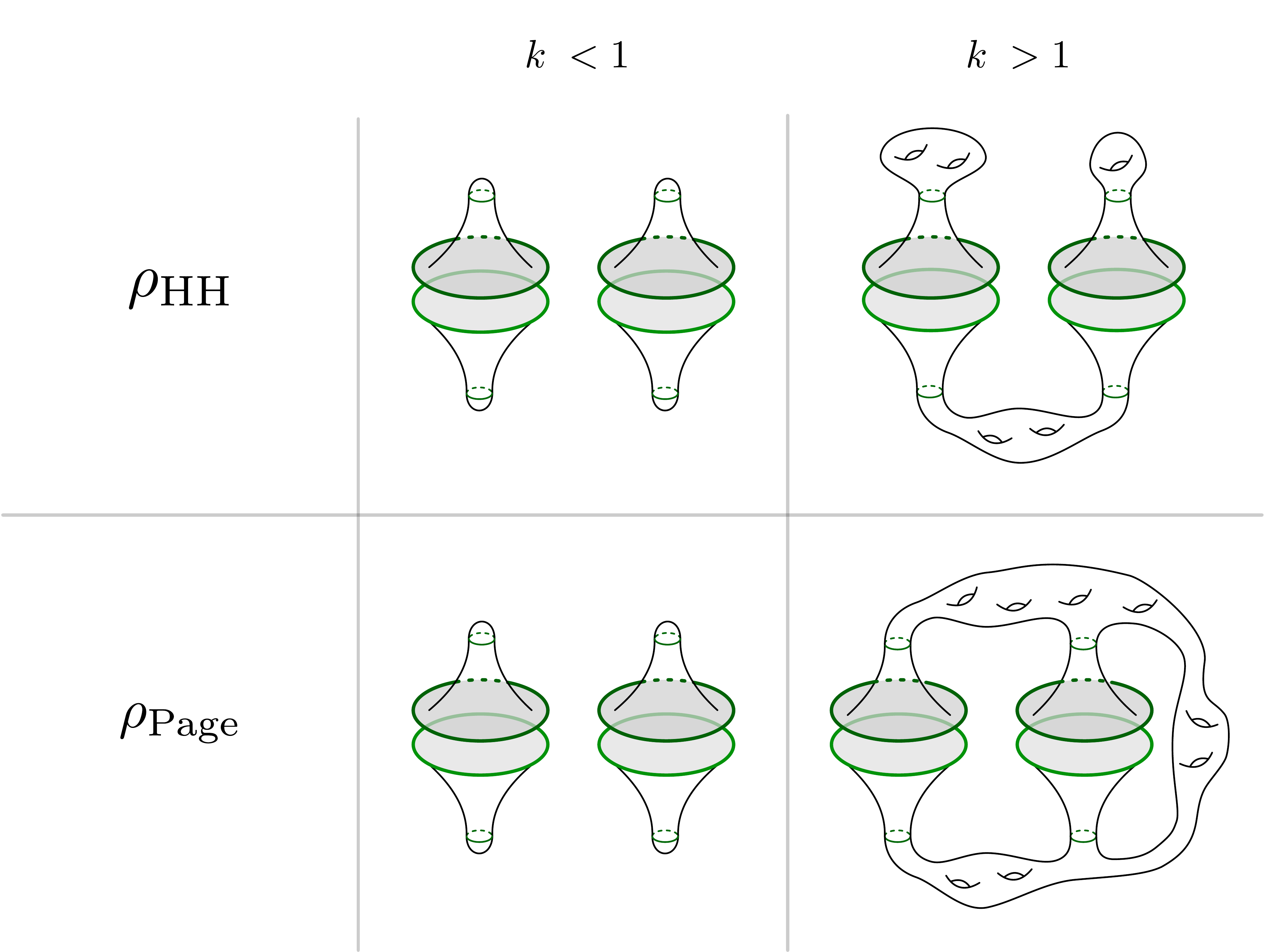}
    \caption{Contributions to two examples of gravitational density matrices. The lighter green boundaries indicate the ket and the darker ones the bra. In the top row we indicated the contributions to the density matrix associated to the Hartle-Hawking state ($\rho_{\rm HH} = \ket{\Psi_{\rm HH}}\bra{\Psi_{\rm HH}}$) for $k < 1$ and $k>1$. In the latter case, any topology other than the disk contributes as long as there are no bra-ket wormholes (wormholes connecting the ket and bra). In the bottom row, we indicated what topologies contribute to the Page density matrix. The situation is similar in the top row; however, since the Page state allows bra-ket wormholes, there can be topologies connecting the bra and ket. Notice that $\rho_{\rm HH} = \rho_{\rm Page}$ for $k<1$, whereas they are unequal for $k>1$ due to the appearance of bra-ket wormholes. In each geometry, we have indicated the gluing of the Lorentzian with the Euclidean section with a small green curve.}
    \label{fig:contributionsrho}
\end{figure}

Now that we understand how to glue higher genus topologies to an asymptotic de Sitter boundary, let us see what fixed $K|_{\partial} = k$\,\footnote{Recall that we only focus on $K> 0$.} and $g_{uu}$ implies. In fact, all conclusions in the AdS case similarly go through in this case.  First, for $k<1$ (as opposed to $k>1$ in AdS$_2$) we only have disk contributions and so this means that when we compute a wavefunction with a number of future boundaries, it will just be the product of the wavefunctions for the disk for each boundary. Furthermore, if we are computing matrix elements of a gravitational density matrix, so there are a number past and future boundaries, there are no bra-ket wormholes, i.e., a geometry connecting the bra and ket boundary, \cite{Chen:2020tes, Anous:2020lka, Penington:2019kki}. Second, for $k>1$, this situation is reversed as there will only be higher genus and no disk contributions. Therefore, matrix elements of the gravitational density matrix would allow for bra-ket wormholes. 

A few comments are in order. As noted in \cite{Anous:2020lka}, when computing gravitational density matrices, it is important to specify the global state in the baby universe Hilbert space $\mathcal{H}_{\rm BU}$. If one computes the Hartle-Hawking density matrix, the matrix elements are computed by a product of two path integrals, and so no bra-ket wormholes can exist in the first place, regardless of what the extrinsic curvatures of the boundaries are. For the Page density matrix, there can be bra-ket wormholes but with these fixed $k$ boundary conditions, they only exist for $k>1$, leading to the non-trivial statement that for $k<1$ the Page density matrix appears to be equivalent to the Hartle-Hawking state. We have summarized the case for the Hartle-Hawking state and Page density matrix in figure \ref{fig:contributionsrho}.

\subsection{$\alpha$-states in JT gravity}

In section \ref{sec:Kguu} and \ref{sec:cosmology}, we saw that depending on whether the extrinsic curvature is bigger or smaller than one, we allow or disallow higher genus corrections to the gravitational path integral. If such higher genus topologies (say in the case of $l$ boundaries with $k>1$) do not contribute, it means that the full perturbative partition function equals the disconnected contribution, i.e., it is a product of $l$ disk partition functions. This allows us to define $\a$-states directly. Following \cite{Marolf:2020xie, Coleman:1988cy, Giddings:1988cx, Giddings:1988wv} we define the operator $\widehat{Z}_{ND}[k,L]$ as an operator in the baby universe Hilbert space $\mathcal{H}_{\rm BU}$ that creates a boundary with proper length $L$ and extrinsic curvature $k$. An $\a$ state is then an eigenstate of the operator $\widehat{Z}_{ND}[k,L]$. 

For an $l$-point function of such boundary creating operators with $k>1$, we have
\be 
\braket{\widehat{Z}_{ND,1}^{>}\cdots \widehat{Z}_{ND,l}^{>}} = \braket{\widehat{Z}_{ND,1}^>}\cdots \braket{\widehat{Z}_{ND,l}^>} = Z^{\rm Disk}_{ND,1}\cdots Z^{\rm Disk}_{ND,l}\,,
\ee
where the subscript indicates the the argument $(k_i,L_i)$, the superscript $k>1$ and the expectation value is in the Hartle-Hawking state. This is a curious equation, because it says that the Hartle-Hawking state is an $\a$ state for $k>1$ boundary creating operators
\be 
\widehat{Z}_{ND}^>[k,L]\ket{\rm HH} = Z_{ND}^{\rm Disk}(k,L) \ket{\rm HH},\quad k>1\,.
\ee
Furthermore, it is easy to construct more $\a$ states $\ket{\a_l}$ by acting with $\widehat{Z}_{ND}^{<}[k,L]$ on the Hartle-Hawking state but now restricting $k<1$,
\be 
\ket{\a_l} = \widehat{Z}_{ND,1}^{<}\cdots \widehat{Z}_{ND,l}^{<}\ket{\rm HH}
\ee
These operators with $k<1$ do not factorize by themselves, since higher genera do contribute in that case, but the $k>1$ operators do again factorize out,
\be 
\braket{\widehat{Z}_{ND,1}^{>}|\a_l} = \braket{\widehat{Z}_{ND,1}^{>}\widehat{Z}_{N,1}^{<}\cdots \widehat{Z}_{ND,l}^{<}} = \braket{\widehat{Z}_{ND,1}^{>}}\braket{\widehat{Z}_{ND,1}^{<}\cdots \widehat{Z}_{ND,l}^{<}}
\ee
so
\be 
\widehat{Z}_{ND,1}^{>}\ket{\a_l} = Z_{ND,1}^{\rm Disk}\ket{\a_l}.
\ee
The $\ket{\a_l}$ states thus constructed are, however, not orthogonal to the Hartle-Hawking state. 

One of the interesting statements about the construction of a Hilbert space using these correlation functions, i.e., the GNS construction, is the existence of null states. From a gravitational perspective, these null states form a highly non-trivial relation between states with a different number of boundaries. For the fixed $k<1$ operators $\widehat{Z}_{ND}[k<1,L]$, the correlators are (up to multiplicative factors) the Weil-Petersson volumes, and the existence of such null states is equivalent to the presence of a highly non-trivial relation between such volumes. Given that such volumes are polynomial in the $L_i$, one could choose coefficients in a superposition of states $\ket{\a_l}$ so that all powers of the $L_i$ have vanishing coefficient once the inner product is taken. It would be interesting to see whether such a system of equations has a non-trivial solution. 

Furthermore, although it is not true that the $\a$-states defined with fixed $K$ can easily give $\a$-states in the theory defined with DD b.c., the reversed statement is true. An $\alpha$-state in the DD theory, which would be the original $\alpha$-states proposed by Marolf and Maxfield, does correspond to an $\alpha$-state in the fixed $K$ theory. That is because, in the DD theory, the $\alpha$-state would be an eigenstate of $\widehat{Z}_{DD}[\phi,L]$ for any $\phi$; thus, the Laplace transform gives an eigenstate of $\widehat{Z}_{ND}[k,L]$, but now for any $k$, not just $k>1$. 

\subsection{Mixed boundary conditions and AdS/CFT}

In the previous discussions, we mostly focussed on fixing either member of a canonical pair. This gave us the four different boundary conditions: DD, ND, DN and NN. In the context of AdS/CFT however, there is a standard way \cite{Aharony:2001pa, Witten:2001ua}\footnote{See also \cite{Papadimitriou:2007sj}.}  to generalise this to mixed boundary conditions, which fixes a certain combination of say, the metric and stress tensor. For a well-defined variational principle, this means that one has to add a multi-trace operator to the boundary field theory, or said differently, the addition of a multi-trace operator is holographically dual to a change in boundary conditions, at least to leading order in large $N$. In \cite{Gross:2019uxi} this was worked out for the case of JT gravity on the disk, assuming a putative dual quantum mechanics. In particular in that case one would start with the JT gravity action with DD boundary conditions, whose variation gives
\be\label{DD}
\d S = \int_{\partial \mathcal{M}} du \sqrt{g_{uu}} \left( \frac{1}{2}T_{uu}\d g^{uu} + O \d \phi \right)\,,
\ee
with (using the Fefferman-Graham gauge \eqref{eq:metricFG})
\be 
T_{uu} = (-2 r g_{uu}(1 - r \partial_r)\phi)|_{\partial},\quad O = (2r^2(1-K))|_{\partial}\,,
\ee
the Hamiltonian density of the dual quantum mechanics and the operator dual to the dilaton, respectively. The variation \eqref{DD} of course vanishes because we have fixed the metric and dilaton on the boundary. From the putative dual quantum mechanics, \eqref{DD} is the usual formula for how the action changes due to a change of the metric and scalar source (in this case the dilaton).

To change the DD boundary conditions, we want to add to the original action a term that can combine into something else that instead of just giving $\d g^{uu}$ or $\d \phi$, could perhaps come from a linear combination that involves $O$ and $T_{uu}$. In \cite{Gross:2019uxi} this idea was employed to determine the term needed to impose the DD boundary conditions at finite radial coordinate (see also \cite{Guica:2019nzm} for the original proposal). In the present context, let us for instance consider adding the following term to the action,
\be \label{dtchange}
S_{\rm d.t.} = \m \int du \sqrt{g_{uu}}O^2
\ee
The variation of this term together with \eqref{DD} can be combined into a variation just like \eqref{DD}, but with $T_{uu}$ and $\phi$ now depending on $\m$,
\be 
\d S_{\rm tot} = \int du \sqrt{g_{uu}}\left( \frac{1}{2} (T_{uu} - \m g_{uu} O^2)\d g^{uu} + O \d(\phi + 2 \m O) \right).
\ee
Thus for a well-defined variational problem from the bulk point of view, we need to fix $\phi + 2 \m O$\,\footnote{Here we work at the classical level and one should think of fixing $\phi + 2 \m O$ as fixing its one-point function.} instead of $\phi$, whereas we still fix $g_{uu}$ on the boundary. Taking $\m$ large we see that we are basically fixing (the one-point function of) $O$, the operator canonically conjugate to the dilaton, which in term of gravitational variables is just $K_r$ if we scale $K = 1 + \e^2 K_r$. The addition of $O^2$ therefore interpolates between DD and ND boundary conditions. 

Let us make two comments regarding this standard way of dealing with different boundary conditions in AdS/CFT. First, the analysis done here is classical and the fate of them at the quantum level is unclear. We have analysed the problem in the $\m = 0$ and $\m \to \infty$ limit in which we could do the computation exactly. However, these two limits are rather special and in particular the ND boundary conditions gave us a topological theory. By tuning $\m$ we thus need to see this transmutation of the theory, which most likely involves a careful analysis of the quantum theory. Furthermore, as we saw in section \ref{sec:Kguu} the sign of $1-K$ is also important, hence the sign of $\m$ will also have to play a crucial role. Second, in contrast to some of the higher dimensional versions of AdS/CFT, in one dimension, we only know of a theory that approximately (there are many massless modes) describes the 2d bulk, namely the Sachdev-Ye-Kitaev (SYK) model. In the IR, this model behaves as a gravitational theory, while in the UV it is a theory of disorderly interacting Majorana fermions. In the IR theory we can identify the boundary value of the dilaton with the inverse of the variance of the gaussian disorder in SYK. The question then remains what is the operator $O$ in SYK model? If we imagine perturbing the dilaton we induce a rescaling of the couplings $J_{ijkl}$ and the change in the action will then be the SYK Hamiltonian. A proposal for the operator $O$ would then simply be the UV Hamiltonian of the SYK model. With this proposal one would then need to check that adding this operator in the UV and flowing to the IR has the desired effect of changing the IR action by \eqref{dtchange}.

\subsection{Boundary Conditions in the Minimal String}

Finally, one can consider an analogous classification of boundary conditions from the perspective of minimal gravity. The connection between JT gravity and the minimal string was first conjectured in \cite{Saad:2019lba}, who obtained the spectral density of JT gravity from the large $p$ limit of the $(2,p)$ minimal string. This connection was elaborated in \cite{Mertens:2020hbs}. This motivates us to extend our classification to this more general setup. We will investigate this connection from an alternate point of view using the Coulomb gas description of the matter field derived in \cite{stanford:unpublished, Kapec:2020xaj} (see also, Appendix F of \cite{Mertens:2020hbs}). The theory considered here may then be viewed as a Coulomb gas CFT ($\chi$) with a gravitational dressing described by a Liouville CFT ($\varphi$) resulting in a Weyl invariant theory with action $S=S_L[\varphi,\hat{g}] + S_M[\chi,\hat{g}]+S_{gh}[b,c,\hat{g}]$,  \footnote{Notice that at the level of the action, one can obtain the matter action from the Liouville action by rotating $b\to i b$ and $\varphi\to -i \chi$ to treat both the theories as Liouville fields $\varphi_+$ and $\varphi_-$ on an equal footing. The central charges are then $c_{\pm}=1+6Q_{\pm}^2=13\pm(b^2+b^{-2})$ and $c_++c_-=26$.} 
\be
S_L[\varphi] &=&  \frac{1}{2}\int_{\mathcal{M}}d^2 x \sqrt{\hat{g}}\left[ (\hat{\nabla}\varphi)^2 + Q\hat{R} \varphi  + 4\pi\mu_L e^{2b\varphi} \right] 
+ \oint_{\partial\mathcal{M}}du \sqrt{\hat{h}}\left[ Q \hat{K}\varphi+ 2\pi \mu_L^B e^{b\varphi} \right] \nonumber \\ 
S_M[\chi] &=& \frac{1}{2} \int_{\mathcal{M}} d^2 x \sqrt{\hat{g}} \left[ -(\hat{\nabla}\chi)^2 - q\hat{R} \chi  + 4\pi\mu_M e^{2b\chi} \right] 
+ \oint_{\partial\mathcal{M}}du\sqrt{\hat{h}}\left[ -q \hat{K}\chi+ 2\pi \mu_M^B e^{b\chi} \right]
\ee
with $\hat{g}$ a fixed background metric, worldsheet coordinates $x$, background charges $Q= b+b^{-1}$, $q= b^{-1}-b$ and central charges $c_L=1+6Q^2$,  $c_M=1-6 q^2$. We have also included the proper boundary terms. The first term is fixed by the Euler character that controls the string expansion and a potential FZZT boundary interaction term with fixed boundary cosmological constant. 
%which is related to the partition function at fixed length via an inverse Laplace transform. 
The parameter $b$ is related to $p$ via $b=\sqrt{\frac{2}{p}}$.

Using the field redefinitions
\be 
\varphi = b^{-1} \rho - b  \Phi,  \qquad\qquad  \chi = b^{-1} \rho +b\Phi,
\ee 
one can establish a correspondence with a `p-deformed' version of JT gravity with a sinh potential for the dilaton \cite{stanford:unpublished, Mertens:2020hbs}
\be \label{eq:minimal-string}
S &=  \int_{\mathcal{M}}d^2 x \sqrt{\hat{g}}\left[ 2\Phi\cdot \hat{\nabla}^2\rho + \hat{R}(\rho- \Phi)  - 4\pi\mu e^{2\rho} \sinh\left(2b^2\Phi\right) \right] \nn\\
&{\,}\,\,\,\,\,  + 2 \oint_{\partial\mathcal{M}}du\sqrt{\hat{h}}\left[  \hat{K}(\rho- \Phi)  - 2\pi\mu^B e^{\rho} \sinh\left(b^2\Phi\right) -\Phi\hat{\partial}_n\rho\right] +S_\text{gh}
\ee
where we also chose $\mu_L=-\mu_M=\mu$ and $\mu_L^B=-\mu_M^B=\mu^B$. 

Let us express the above action in terms of a physical (dynamical) metric $g$ which is related to the fiducial metric $\hat{g}$ via a conformal factor which we identify with the field $\rho$, namely $g_{\mu\nu} = e^{2\rho} \hat{g}_{\mu\nu}$. To express the action in terms of the physical metric, we use the Weyl transformation properties, 
 \be 
 R=e^{-2\rho}(\hat{R}-2\hat{\nabla}^2\rho),\qquad \qquad K=e^{-\rho}\left( \hat{K} +  \hat{\partial}_n \rho \right),
\ee 
where hatted quantities are evaluated in the background metric. The action is then given by 
\be 
S &=  -\int_{\mathcal{M}}d^2 x \sqrt{g}\left[ \Phi R - R\rho +2(\nabla\rho)^2   + 4\pi\mu  \sinh\left(2b^2\Phi\right) \right] \nn\\
&{\,}\,\,\,\,\,  - 2 \oint_{\partial\mathcal{M}}du\sqrt{h}\left[  K\Phi -K\rho  + 2\pi\mu^B \sinh\left(b^2\Phi\right) \right] +S_\text{gh}\,.
\ee  

In the $p\to\infty$ limit, one recovers ordinary JT with the identification of the physical metric $g_{\mu\nu}=e^{2\rho}\hat{g}_{\mu\nu}$ in terms of the fiducial metric $\hat{g}$  and by scaling the cosmological constant to large values such that $4\pi\mu b^2=\Lambda_{JT}=1$ and $2\pi\mu^B b^2=1$ .  In the strict JT limit, the conformal mode $\rho$ is non-dynamical and kinetic term involving solely $\rho$ may be omitted as it contributes to an overall constant in the path integral
\footnote{The same is also true for the ghost part of the action $S_{gh} =  \int d^2z \sqrt{\hat{g}} (b\hat{\nabla} c +\bar{b}\hat{\nabla}\bar{c})$,  which does not depend on $\rho$ and $\Phi$. }
\be \label{eq:minimal-jt}
S \stackrel{b\to 0}{\approx} -\int_{\mathcal{M}}d^2 x \sqrt{g}\left[\Phi(R+2)\right]-2\oint_{\partial\mathcal{M}}du\sqrt{h}\left[ \Phi (K-1)  \right].
\ee 
Hence, we see the correspondence between the degrees of freedom in the two descriptions
\be \label{eq:minimal-correspondence1}
(e^{2\rho})_\text{Liouville}\leftrightarrow (g_{uu})_\text{JT}, \qquad\qquad (\Phi)_\text{Liouville}\leftrightarrow (\phi)_\text{JT}.
\ee 
where all the quantities refer to their boundary values. 

To classify the boundary conditions, let us compute the variation of the action 
\be 
\delta S &= \text{EOM} -2 \oint_{\partial\mathcal{M}}du\sqrt{\hat{h}}\left(\delta\Phi (\hat{\partial}_n\rho+\hat{K})  + \delta\rho (\hat{\partial}_n\Phi-\hat{K})+2\pi\mu^B\delta\left( e^{\rho}\sinh(b^2\Phi) \right)\right)  \nn \\
%&= \text{EOM} -2 \oint_{\partial\Sigma}du\sqrt{h}\left(\delta\Phi K  + \delta\rho (\partial_n\Phi-K+\partial_n\rho)\right)  \nn \\
&= \text{EOM} -2 \oint_{\partial\mathcal{M}}du\left[\delta\Phi e^{\rho}(K+2\pi\mu^B b^2\cosh(b^2\Phi)) \right.  \nn \\
&{ }\hspace{1.5 in} \left. + \delta(e^{\rho}) \left(\partial_n\Phi-K+\partial_n\rho+2\pi\mu^B\sinh(b^2\Phi)\right) \right] . 
\ee
where in the last equation, we took $h_{\mu\nu}dx^{\mu}dx^{\nu} = e^{2\rho} du^2$. We see that the variation vanishes when fixing $\rho$ and $\Phi$ on the boundary. This is analogous to the DD boundary condition in JT gravity via the identification \eqref{eq:minimal-correspondence1} and is consistent with the corresponding boundary term appearing in \eqref{eq:minimal-jt}. 

Hence, we see that the alternate boundary conditions involve fixing the energy in this theory or fixing the extrinsic curvature in the physical metric instead of $\phi$ and $\rho$ via a Legendre transform to the corresponding conjugate variables. So to complete the mapping, we have 
\be 
(K+2\pi\mu^B b^2\cosh(b^2\Phi))_\text{Liouville}&\leftrightarrow (K)_\text{JT}, \nn \\ (\partial_n\Phi-K+\partial_n\rho+2\pi\mu^B\sinh(b^2\Phi))_\text{Liouville}&\leftrightarrow (\partial_n\phi-\phi K)_\text{JT}\,,
\ee 
which leads to an analogous four-fold classification of boundary terms. We see that the conjugate variables are $p$-deformed by the boundary interaction. In addition, we also have extra contributions from the variation of the $\rho$ mode that we have dropped earlier in the JT limit. This implies that one must be wary of the fact that the operations of varying the action and taking the JT limit do not commute.

So far, the analysis is exact at the level of the path integral. However, the boundary conditions implied here only match in the  $b\to 0$ limit when compared to those used in \cite{Mertens:2020hbs}. 
This is related to the choice of normal ordering when quantizing this theory. It would be interesting to study the full quantum mechanical partition function and amplitudes in this theory and its matrix model dual. It would also be interesting to see if this theory can be directly related to the gas of defects studied recently in \cite{Maxfield:2020ale, Witten:2020wvy,Alishahiha:2020jko}.

\subsection{Future directions}

To conclude, we discuss a number of possible future directions that would make use of the results derived in this paper.

\subsubsection*{Higher dimensions}

Thus far, we have only considered two-dimensional dilaton gravities, but most of our analysis could also be applied to higher-dimensional gravity theories, albeit now with $g_{\mu\nu}$ and $K_{\mu\nu}$ as phase space variables. The DD and DN are relatively straightforward as they map onto the canonical and microcanonical ensembles in the higher dimensional field theory \cite{PhysRevD.47.1420, Compere:2008us}; however, it would be interesting to figure out the higher-dimensional analogs of the ND and NN boundary conditions considered in this paper.\footnote{Fixing the trace of the extrinsic curvature in higher dimensional gravity appears to be a better behaved boundary condition (together with fixing the conformal class of the metric), as reviewed in \cite{Witten:2018lgb}.} In fact, \cite{Belin:2020oib} proposed an explicit construction using the higher dimensional analogue of the $T\bar{T}$ deformation for studying $ND$ boundary conditions along a spacelike boundary.

A particularly tractable theory of gravity that shares many features with the theory of gravity considered here is three-dimensional gravity in its Chern-Simons formulation. In this case, the gauge algebra is $\mathfrak{sl}(2,\mR) \times \mathfrak{sl}(2,\mR)$ and gives rise to two gauge fields $A$ and $\,\bar{A}$. We can consider various boundary conditions on these gauge fields that in the second order formulation correspond to fixing certain components of the metric and extrinsic curvature. The fixed ND boundary condition would be one where the Chern-Simons theory does not have any boundary terms and is purely topological. It would therefore be interesting to study the exact computation of the partition functions in that case.  

\subsubsection*{Finite cutoff JT}

Recently there has been a great effort in trying to define two-dimensional gravity for finite patches of space-time \cite{Iliesiu:2020zld, Gross:2019ach, Gross:2019uxi}.\footnote{See also, \cite{McGough:2016lol} for the original proposal with a three-dimensional bulk, or \cite{Hartman:2018tkw, Taylor:2018xcy} for higher-dimensional analogs.} In doing so, one deforms the putative quantum mechanics dual with a particular operator that depends solely on the Hamiltonian and coupling constant. This is a well-known story in the standard Dirichlet case, but it would be interesting to determine whether other boundary conditions can also be insightful for defining gravity at finite cutoff. For instance, the spectrum of the deformed quantum mechanics theory is known to exhibit a complexification of the energy levels at high energy,\footnote{See \cite{Stanford:2020qhm} for a entirely different approach that did not yield such a complexification.} signaling the presence of a UV cutoff. However, upon Laplace transforming in $\phi$ (functionally), we arrive at the ND boundary conditions, whose partition function is simple and topological. It would be interesting to use the simplicity of the ND partition functions, say for the disk, to study the DD finite cutoff theory in more detail. In particular, using the techniques above, we can, in principle, determine what happens when the cutoff surface is brought inside all the way to the center of the disk, where we expect it to reproduce the Bekenstein-Hawking entropy. Putting the DN boundary conditions (the microcanonical ensemble) can also be considered at finite cutoff, which we hope could shed more light on the complexification of the previously discussed energies. See also \cite{Coleman:2020jte} for a discussion of these ideas in three dimensions.

\subsubsection*{Non-Perturbative Corrections}

When studying the ND boundary conditions, we have seen that all perturbative corrections in $e^{\phi_0}$ vanish provided that $k>1$. It is then natural to wonder if this statement is also true non-perturbatively for a UV completion of JT gravity, such as its matrix integral completion.\footnote{Alternatively, requiring that all non-perturbative corrections vanish for $k>1$ can be viewed as a condition on the possible UV completions of JT gravity.  } In particular, we expect that the kernel presented in section \ref{sec:ND-matrix-integral-interp}, when integrated against non-perturbative terms in the partition function, would kill these terms in addition to the perturbative terms. If the non-perturbative completion is specified by a matrix model, this translates to the expectation value of the operator \eqref{eq:ND-matrix-operator} vanishing.

It would be interesting to use the results of \cite{Saad:2019lba} to study a non-perturbative completion of JT gravity via the corresponding double-scaled matrix model to check if this is true. A complete understanding of this requires detailed knowledge of such non-perturbative corrections, including their exact form at exponentially small energies. Nevertheless, one cannot rule out the possibility that a careful application of the integration contour for the kernel, together with some universal analytic properties of the non-perturbative contributions, can indeed lead to this conclusion. 

It would also be interesting to carry out a similar analysis for the supersymmetric cousins of JT gravity studied in \cite{Stanford:2019vob} and obtain the analog of our kernel, specifically for the $\left(\alpha=\{0,1,2\},\beta=2\right)$ Altland-Zirnbauer ensembles. The advantage of these JT supergravity models is that their non-perturbative behavior is under much better control than in standard JT. It is interesting to note that these theories exhibit a truncation in the perturbative series, even with standard boundary conditions in some cases. Hence, they are better suited to study factorization properties, $\alpha$-branes, and non-perturbative phenomena. We leave such an analysis to future work.

\section*{Acknowledgment}

We are grateful for discussions with Victor Gorbenko, Sean Hartnoll, Douglas Stanford, Joaquin Gustavo Turiaci, and Mark Mezei. A special thanks goes to Raghu Mahajan for valuable  comments about the draft. JK and ZY is supported by the Simons Foundation. LVI was supported in part by the US NSF under Grant No. PHY-1820651, by the Simons Collaboration on the Conformal Bootstrap, a Simons Foundation Grant with No. 488653 and by the Simons Collaboration on Ultra-Quantum Matter, a Simons Foundation Grant with No. 651440.

\appendix 
\section{The variation of the bulk action}\label{app:variationbulkaction}

In this appendix we compute the variation of the bulk action 
\be
S_{\rm bulk} = - \frac{1}{2\k^2} \int_{\mathcal{M}} d^2 x \sqrt{g}\, (\phi R - 2 U(\phi))
\ee
In particular, we will focus on the boundary terms, which are of prime importance in the bulk of the text. We will set $2\k^2 = 1$. Abstractly the variation will take the form 
\be 
\d S_{\rm bulk} = -\int_{\mathcal{M}} d^2 x \sqrt{g} \left( E_{\phi} \d \phi + E_{\mu\nu} \d g^{\mu\nu} + \nabla^{\mu} \Theta_{\mu} \right)
\ee
The first two terms give the equations of motion: 
\be
E_{\phi} = (R + 2),\quad  E_{\mu\nu} = \left(\nabla^\mu \nabla^\nu - g^{\mu\nu}\nabla^2\right)\phi - g_{\mu\nu}U(\phi), 
\ee
whereas $\nabla^\mu \Theta_{\mu}$ gives rise to a boundary term in the variation. To derive the equations of motion and an explicit form for $\Theta_{\mu}$, we use
\be 
\d R = R_{\mu\nu} \d g^{\mu\nu} + \nabla^\rho (g^{\mu\nu} \d\G^{\rho}_{\nu \mu} - g^{\mu\rho}\d\G^{\nu}_{\nu\mu}),
\ee
with $\G$ the Christoffel symbol associated to $g$. Inserting its explicit form, we can write the term in brackets as 
\be 
g^{\mu\nu} \d\G^{\rho}_{\nu \mu} - g^{\mu\rho}\d\G^{\nu}_{\nu\mu} = g^{\nu\rho} g^{\a\b} (\nabla_\beta \d g_{\a\nu} - \nabla_\nu \d g_{\a\b})
\ee
and the variation of the bulk action becomes 
\begin{align}
\d S_{\rm bulk} &= - \int_{\mathcal{M}} d^2 x \sqrt{g} \left( \d \phi (R-2U'(\phi))  - \nabla^\rho \phi ( \nabla^\a \d g_{\a\rho} - g^{\a\b}\nabla_\rho \d g_{\a\b}) \right.\nonumber\\
&\left. \quad -  U(\phi)g^{\mu\nu}\d g_{\mu\nu} \right) - \int_{\partial \mathcal{M}} du \sqrt{h} \phi n^{\mu} (\nabla^\a \d g_{\a\mu} - g^{\a\b}\nabla_{\mu} \d g_{\a\b}),
\end{align}
with $h$ the induced metric on the boundary. Notice that in $2d$ we have $R_{\mu\nu} = \frac{1}{2}R g_{\mu\nu}$, $K_{\mu\nu} = K h_{\mu\nu} = \nabla_\a n^\a h_{\mu\nu}$ and 
\be
n^{\mu} (\nabla^\a \d g_{\a\mu} - g^{\a\b}\nabla_{\mu} \d g_{\a\b}) = - 2 \d K + D_{\nu} (n^{\nu}n^{\a}n^\b \d g_{\a\b}) - 2 K^{\mu\nu}\d g_{\mu\nu} - n^\nu h^{\a\b}\nabla_{\a}\d g_{\b\nu}
\ee
Here $D_{\nu}=h_{\mu\nu}\nabla^{\mu}$ is the boundary covariant derivative. After doing a couple of partial integrations it is straightforward to verify that the equations are as given above. The boundary term takes the form
\be 
\delta S_{\rm bulk,\partial} = - \int_{\partial \mathcal{M}} \sqrt{h} du \left[ \left(\partial_n \phi - K \phi\right) g^{\a\b} \d g_{\a\b} - 2 \phi \d K \right]
\ee
This term needs to cancelled by an appropriated boundary term depending on what boundary conditions are chosen in order to have a well-defined variational problem.

\section{Schwarzian calculation for fixed $K$ and $g_{uu}$}
\label{app:SchwarzianFixedK}

Here we will perform a direct Schwarzian calculation of the fixed $K$ boundary path integral in the asymptotic limit.
Let's start with the disk case first, the fixed $K$ boundary condition can be obtained from the usual dirichlet boundary condition by doing a functional integral over $\phi_b={\phi_r\over \varepsilon}$, this gives us the following boundary integral:
\be 
Z_{ND} &= \int D\phi_b(u) e^{-2 \int_0^\beta du \sqrt{g_{uu}}\phi_b(u) k}Z_{DD}(\phi_b(u)) = \int  D\phi_r(u) \int D f(u) e^{\int_0^\b du \phi_r(u) (\Sch(f, u) - \kappa)}\nn \\  &=  
\int D f(u) \delta (\Sch(f,u)-\kappa), 
\ee
where $\kappa={k-1\over \varepsilon^2}$ is the regularized extrinsic curvature in the asymptotic limit.
This integral can be done by perturbation theory, where we expand 
\be
f(u)=\tan\left({\pi\over \beta}(u+\epsilon(u))\right),\quad \epsilon(u)=\sum_{|n|\geq 2} e^{-{2\pi\over \beta} i n u}(\epsilon^{(R)}_n+i\epsilon^{(I)}_n),
\ee
and we have $\epsilon^{(R)}_n=\epsilon^{(R)}_{-n}$ and $\epsilon^{(I)}_n=-\epsilon^{(I)}_{-n}$.  The integral measure for $\epsilon$ is determined by their symplectic form \cite{Saad:2019lba}
\be
\Omega=2 {(2\pi)^3\over \beta^2}\sum_{n\geq 2}(n^3-n)d\epsilon_n^{(R)}\wedge d\epsilon^{(I)}_n.
\ee
To the first order expansion of $\epsilon$, the delta function of Schwarzian function is equal to
\be
\begin{split}
&\delta \left({2\pi^2\over \beta^2}-\kappa+\sum_{n\geq 2} 2{(2\pi)^3\over \beta^3} (n^3-n)(\epsilon_n^{(R)}\sin {2\pi n u\over \beta}-\epsilon_n^{(I)}\cos{2\pi n u\over \beta}) \right)\\
&=\delta\left({2\pi^2\over \beta^2}-\kappa\right)\delta(0)^2 \prod_{n\geq 2}{1\over 2 {(2\pi)^6\over \beta^6} (n^3-n)^2}\delta(\epsilon_n^{(R)})\delta(\epsilon_n^{(I)}),
\end{split}
\ee
where the additional two $\delta(0)$s come from the $n=\pm 1$ fourier modes.
It is then straightforward to direct evaluate the path integral
\be
Z_{ND}^{(\text{Disk})}=\delta\left({2\pi^2\over \beta^2}-\kappa\right)\delta(0)^2\prod_{n\geq 2} {\beta^4\over (2\pi)^3} {1\over n^3-n}=\delta\left({2\pi^2\over \beta^2}-\kappa\right)\delta^2(0),
\ee
where in the last equality we absorbed the regularized product (which is finite) in $\delta^2(0)$. 
Let's now look at the trumpet partition funciton with geodesic length $b$. 
The perturbation of the boundary wiggle is
\be
f(u)=e^{-{b\over \beta}(u+\epsilon(u))};~~~\epsilon(u)=\sum_{|n|\geq 1} e^{-{2\pi\over \beta} i n u}(\epsilon^{(R)}_n+i\epsilon^{(I)}_n).
\ee
The symplectic measure for $\epsilon$ is now
\be
\Omega=2 {(2\pi)^3\over \beta^2}\sum_{n\geq 1}\left(n^3+{b^2\over (2\pi)^2}n\right)d\epsilon_n^{(R)}\wedge d\epsilon^{(I)}_n.
\ee
The delta function of the Schwarzian variable to linear order in $\epsilon$ is equal to
\be
\begin{split}
&\delta \left(-{b^2\over 2\beta^2}-\kappa+\sum_{n\geq 1} 2{(2\pi)^3\over \beta^3} (n^3+{b^2\over (2\pi)^2}n)(\epsilon_n^{(R)}\sin {2\pi n u\over \beta}-\epsilon_n^{(I)}\cos{2\pi n u\over \beta}) \right)\\
&=\delta\left({b^2\over 2\beta^2}+\kappa\right)\prod_{n\geq 1}{1\over 2 {(2\pi)^6\over \beta^6} \left(n^3+{b^2\over (2\pi)^2}n\right)^2}\delta(\epsilon_n^{(R)})\delta(\epsilon_n^{(I)}).
\end{split}
\ee
Putting these together, the single trumpet partition function is given by
\be
Z_{ND}^{(\text{Trumpet})}=\delta\left({b^2\over 2\beta^2}+\kappa\right)\prod_{n\geq 1} {\beta^4\over (2\pi)^3 }{1\over n^3+{b^2\over (2\pi)^2}n}=\delta\left({b^2\over 2\beta^2}+\kappa\right){b\over 2 \beta^2 \sinh{b\over 2}},
\ee
where we use the formula
\be
\sum_{n\geq 1}\log( n^2+a^2)=\log {2\sinh a\pi\over a}.
\ee
The cylinder result can be obtained by gluing two trumpets with the WP measure of $b$, this give us
\be
\int b db  Z_{ND}^{(\text{Trumpet})}(\beta_1,\kappa_1;b)Z_{ND}^{(\text{Trumpet})}(\beta_2,\kappa_2;b)={-2 \kappa_1\beta_1^2\over  \sinh^2\left(\beta_1\sqrt{-\kappa_1\over 2}\right)}\delta(\kappa_1\beta_1^2-\kappa_2\beta_2^2).
\ee
To make comparison with formula \ref{eq:cylinder} and \ref{eq:ND-trumpet}, we can define the holonomy $\lambda_{1,2}=\beta_{1,2}\sqrt{-2\kappa_{1,2}}$.
Then the cylinder partition function and trumpet partition function can be rewritten as:
\be
Z_{ND}^{(\text{Cylinder})}={\lambda_1\over  4\sinh^2{\lambda_1\over 2}}\delta(\lambda_1-\lambda_2),\quad Z_{ND}^{(\text{Trumpet})}={\delta(b-\lambda)\over 2\sinh{b\over 2}}.
\ee

\section{Delta functions and the Weyl integration formula}
\label{sec:delta-functions-app}

Here we consider the different delta functions that appeared in the main text in more detail. We will first consider the compact group case, exemplified with $SU(2)$ and move on to $PSL(2,\mathbb{R})$ afterwards. These discussions are just to gain some intuition. The more rigorous unifying framework to discuss both cases at once is the Weyl integration formula, which we discuss at the end. 

\subsection{The case for $SU(2)$}

As is well known, the delta function $\d(g)$ on a group manifold $G$ for $g \in G$ has two different interpretations. On the one hand, we can think of the Lie group $G$ as a manifold and locally one can pick a metric with which one can define the delta function in the usual sense. For instance, consider the group $SU(2)$. This Lie group is isomorphic to the $3$-sphere. A metric on this three sphere is
\be 
ds^2 = d\a^2 + \sin^2 \a \;d\b^2 + \sin^2 \a \sin^2 \b \;d\g^2,\quad \a,\b \in (0,\pi],\g \in [0,2\pi),
\ee
with measure $d^3 x = \sin^2 \a \sin \b\;d\a d\b d\g$. In these coordinates $SU(2)$ group elements are parametrized as
\be 
g = \begin{pmatrix}
\cos \a + i \cos \b \sin\a & e^{i\g} \sin \a \sin \b \\
-e^{-i\g} \sin \a \sin \b & \cos \a - i \cos \b \sin\a
\end{pmatrix}
\ee
The delta function on $SU(2)$ is
\be 
\d_{SU(2)}(g) = \frac{1}{\sin^2 \a \sin \b} \d(\a)\d(\b)\d(\g).
\ee
This delta function is then defined for smooth test functions $f$ on the three sphere or equivalently on $SU(2)$. Suppose now that we restrict the space of test functions to only those that depend on the conjugacy classes of $SU(2)$, which with our parametrisation are parametrized by $\a$. So we have $f = f(\a)$. On the sphere, these are spherical symmetric test functions. For such functions, the delta function is slightly different as can be seen as follows,
\be 
f(0) = \int dg \d_{SU(2)}(g)f(g) = \int d^3 x \frac{1}{\sin^2 \a \sin \b} \d(\a)\d(\b)\d(\g) f(\a) = \int_0^\pi d\a \sin^2 \a f(\a) \frac{\d(\a)}{\sin^2 \a} 
\ee
So on the 'spherically symmetric' test functions, the delta function is 
\be 
\d_c(g) = \frac{\d(\a)}{\sin^2\a}
\ee
with the subscript $c$ referring to 'conjugacy class'. Yet another way of thinking about the delta function $\d(g)$ is through the character decomposition:
\be
\d(g) = \sum_{R} \dim R \; \chi_R(g)
\ee
where the sum is over all irreducible representations of $G$. From this expression, it is clear thet $\d(g)$ is trace class. For $G = SU(2)$ this gives 
\be 
\d(g) = \sum_{l=1}^{\infty} l\;\frac{\sin l \a}{\sin \a} = -\pi \frac{\d'(\a)}{\sin \a}.
\ee
This seems different from what we had obtained previously, but a closer look will reveal they are the same, up to normalisation. Again, we need to integrate this delta function against test functions that only depend on the conjugacy class of $g$, i.e. only depend on $\a$ and the measure is important. From the previous calculation, we know the measure is $\sin^2 \a$, so 
\be 
\int_0^{\pi} -\pi \frac{\d'(\a)}{\sin \a} f(\a) \sin^2 \a = \pi f(0)\,,
\ee
and hence, upto normalisation this agrees with $\delta_c(g)$.

\subsection{The case for $PSL(2,\mR)$}

Let us now consider another example that we focussed on in the main text: $G = PSL(2,\mathbb{R})$. The manifold for this group is AdS$_3$ with metric
\be 
ds^2 = da^2 - \sinh^2 a\;db^2 + \sinh^2 a \cosh^2 b\;dc^2
\ee
with $a, b$ run over the entire real line and $c$ is between $0$ and $2\pi$. With this parametrisation we can represent hyperbolic elements of $SL(2,\mathbb{R})$ as
\be 
g = \begin{pmatrix}
\cosh a + \cos c \cosh b \sinh a && \sinh \a (\cosh b \sin c + \sinh b)\\
\sinh \a (\cosh b \sin c - \sinh b) && \cosh a - \cos c \cosh b \sinh a 
\end{pmatrix}
\ee
In these elements $a$ parametrizes the conjugacy class. Let us consider test functions that are just functions of $a$. The delta function on such functional spaces is then given by 
\be 
\d_c(g) = \frac{\delta(a)}{\sinh^2 a}
\ee
Here is a subtle difference between $SU(2)$ and $SL(2)$, in $SU(2)$ there is just elliptic elements, whereas for $PSL(2,\mathbb{R})$ we have two more conjugacy classes, so the element $g$ above only covers the hyperbolic elements. This is saying the above coordinates are not global AdS$_3$ coordinates. To understand the general case for $PSL(2,\mathbb{R})$ we invoke Weyl's integration formula.

\subsection{Weyl integration formula}

Weyl's integration formula is a formula that relates the integral of a function on the group integrated over the entire group $G$ to the integral of a slightly different function over the Cartan subgroups $H_i$, $i = 1,\cdots r$. Let $f(g)$ be a compactly supported function on $G$, then Weyl's integration formula is the following statement, (see \cite{knapp2001representation}, proposition 5.27 for more details)
\be 
\int_G dg f(g) = \sum_{i = 1}^r \frac{1}{|W_{H_i}|} \int_{H_i} dU |D_{H_i}(U)|^2 \left( \int_{G/H_i} d\hat{g} f(\hat{g} g \hat{g}^{-1}) \right).
\ee
Lets unpack this formula. Here we assumed the group $G$ having $r$ Cartan subgroups with Weyl group $W_{H_i}$. The term in round brackets is an integral over the quotient $G/H_i$. The most important term here is $D_{H_i}(H)$, which explicitly reads
\be \label{Wdenom}
D_{H_i}(U) = \chi_{\d}(U) \prod_{\a \in \D^+} (1 - \chi_{\a}(U)^{-1}) = \prod_{\a \in \D^+} (\chi_{\a/2}(U) - \chi_{\a/2}(U)^{-1}),
\ee
where $\chi$ is the character of the group $H_i$ with the subscript indicating the representation of the Lie algebra of $H_i$. Explicitly, for the cases we will consider the $H_i$ will be abelian and the characters are just exponentials,
\be 
\chi_{\a}(U) = e^{\a(u)}
\ee
with $u$ the corresponding Lie algebra element to $U$. The product in \eqref{Wdenom} is over all positive roots $\a$ and $\d$ half the sum of all positive roots. 

Let us now consider out two examples $PSL(2,\mathbb{R})$ and $SU(2)$. In the latter case, there is one Cartan subgroup $H$ given by elements $\diag(e^{i\theta}, e^{-i \theta})$. The roots system is well known and there is only a single positive root. We have $\chi_{\a/2}(U) = e^{i \theta}$. So,
\be 
|D_{H}(U)|^2 = |e^{i\theta} - e^{-i\theta}|^2 = 4 \sin^2 (\theta)
\ee
Furthermore $W_{H} = 2$. If we now consider a trace-class function $f$ we just get the volume of $G/H$, which is $4\pi$. So we get, 
\be 
\int_{SU(2)} dg f(g) = \frac{1}{2} \int_0^{2\pi} \frac{d\theta}{2\pi}\;4 \sin^2(\theta) f(\theta),
\ee
where we normalized the Haar measure so that $SU(2)$ has volume $1$. 

Let us now move on to $PSL(2,\mathbb{R})$. In this case there are two non-conjugate Cartan subgroups, whose algebra is generated by the Pauli matrices $i\sigma_2$ and $\sigma_3$. Let us denote these groups by $H_1$ and $H_2$. For the former algebra the situation is the same as the previous discussion, but for the Cartan algebra generated by $\s_3$, the characters are real: $\chi_\a(U) = e^{a}$ with $U = e^{a \s_3/2}$. Furthermore, we have $|W_{H_1}| = 1$ and $|W_{H_2}| = 2$. Considering a trace-class function $f$ we get
\be 
\int_{PSL(2,\mathbb{R})} dg f(g) = \vol\,G \left(\int_0^{2\pi} \frac{d\theta}{2\pi}\;4 \sin^2(\theta) f(\theta) + \frac{1}{2} \int_0^{\infty} \frac{da}{\vol\,H_2}\;4 \sinh^2(a/2) f(a)\right).
\ee
Because we took here a function that is trace-class, the integrals diverge and we need to regularize them. For non-trace-class functions, such divergences will not occur as long as the functions have compact support.

\bibliographystyle{ssg}
\bibliography{Biblio}

\end{document}